\documentclass[useAMS,usenatbib]{mn2e}
\usepackage{psfig}
\usepackage{graphicx}

\title[Cooling flows, black holes and the luminosities and colours of galaxies]
{The many lives of AGN: cooling flows, black holes and the luminosities
and colours of galaxies}

\author[Croton et al.]{
\parbox[t]{\textwidth}{
Darren J. Croton$^{1}$,
Volker Springel$^{1}$.
Simon D. M. White$^{1}$,
G. De Lucia$^1$,
C. S. Frenk$^2$,
L. Gao$^1$,
A. Jenkins$^2$,
G. Kauffmann$^1$,
J. F. Navarro$^3$,
N. Yoshida$^4$
}
\vspace*{6pt} \\ 
$^1$Max-Planck-Institut f\"ur Astrophysik, D-85740 Garching, Germany.\\
$^2$Institute for Computational Cosmology, Physics Department, Durham. U.K.\\
$^3$Department of Physics and Astronomy, University of Victoria, B.C., Canada.\\
$^4$Department of Physics, Nagoya University, Chikusa-ku, Nagoya 464-8602, Japan\\
\vspace{-0.5cm} 
}

\date{Accepted ---. Received ---;in original form ---}

\pubyear{2005}

\newcommand{\plotone}[1]
           {\centering \leavevmode \psfig{file=#1,width=\columnwidth,clip=}}

\newcommand{\plotfull}[1]
           {\centering \leavevmode \psfig{file=#1,width=\textwidth,clip=}}

\begin{document}
\def\simlt{\lower.5ex\hbox{$\; \buildrel < \over \sim \;$}}
\def\simgt{\lower.5ex\hbox{$\; \buildrel > \over \sim \;$}}

\maketitle


\begin{abstract}
We simulate the growth of galaxies and their central supermassive
black holes by implementing a suite of semi-analytic models on the
output of the Millennium Run, a very large simulation of the
concordance $\Lambda$CDM cosmogony. Our procedures follow the detailed
assembly history of each object and are able to track the evolution of
all galaxies more massive than the Small Magellanic Cloud throughout a
volume comparable to that of large modern redshift surveys. In this
first paper we supplement previous treatments of the growth and
activity of central black holes with a new model for `radio' feedback
from those AGN that lie at the centre of a quasistatic X-ray emitting
atmosphere in a galaxy group or cluster.  We show that for
energetically and observationally plausible parameters such a model
can simultaneously explain: (i) the low observed mass drop-out rate in
cooling flows; (ii) the exponential cut-off at the bright end of the
galaxy luminosity function; and (iii) the fact that the most massive
galaxies tend to be bulge-dominated systems in clusters and to contain
systematically older stars than lower mass galaxies.  This success
occurs because static hot atmospheres form only in the most massive
structures, and radio feedback (in contrast, for example, to supernova
or starburst feedback) can suppress further cooling and star formation
without itself requiring star formation. We discuss possible physical
models which might explain the accretion rate scalings required for
our phenomenological `radio mode' model to be successful.
\end{abstract}


\section{Introduction}
\label{intro}
The remarkable agreement between recent measurements of cosmic
structure over a wide range of length- and time-scales has established
a standard paradigm for structure formation, the $\Lambda$CDM
cosmogony. This model can simultaneously match the microwave
background fluctuations seen at $z\!\sim\!  1000$
\citep[e.g.][]{Spergel2003}, the power spectrum of the low redshift
galaxy distribution \citep[e.g.][]{Percival2002, Tegmark2004}, the
nonlinear mass distribution at low redshift as characterised by cosmic
shear \citep[e.g.][]{vanWaerbeke2002} and the structure seen in the
$z\!=\!3$ Ly~$\alpha$ forest \citep[e.g.][]{Mandelbaum2003}. It also
reproduces the present acceleration of the cosmic expansion inferred
from supernova observations \citep{Perlmutter1999, Riess1998}, and it
is consistent with the mass budget inferred for the present universe
from the dynamics of large-scale structure \citep{Peacock2001}, the
baryon fraction in rich clusters \citep{White1993} and the theory of
Big Bang nucleosynthesis \citep{Olive2000}. A working model for
the growth of all structure thus appears well established.

In this cosmogony, galaxies form when gas condenses at the centres of
a hierarchically merging population of dark haloes, as originally
proposed by \cite{White1978}. Attempts to understand this process in
detail have consistently run into problems stemming from a mismatch in
shape between the predicted distribution of dark halo masses and the
observed distribution of galaxy luminosities.  Most stars are in
galaxies of Milky Way brightness; the galaxy abundance declines
exponentially at brighter luminosities and increases sufficiently
slowly at fainter luminosities that relatively few stars are in
dwarfs. In contrast, the theory predicts a much broader halo mass
distribution -- a constant mass-to-light ratio would produce more high
and low luminosity galaxies than are observed while underpredicting
the number of galaxies like the Milky Way. Attempts to solve these
problems initially invoked cooling inefficiencies to reduce gas
condensation in massive systems, and supernova feedback to reduce star
formation efficiency in low mass systems \citep{White1978,
White1991}. Formation of dwarfs may also be suppressed by
photoionisation heating \citep{Efstathiou1992}. As \cite{Thoul1995}
emphasised, cooling effects alone are too weak to produce the bright
end cut-off of the luminosity function, and recent attempts to fit
observed luminosity functions have been forced to include additional
feedback processes in massive systems \citep[e.g.][]{Benson2003}.  In
this paper we argue that radio sources may provide the required
feedback while at the same time providing a solution to two other
long-standing puzzles.

An important unanswered question is why the gas at the centre of most
galaxy clusters is apparently not condensing and turning into stars
when the observed X-ray emission implies a cooling time which is much
shorter than the age of the system. This cooling flow puzzle was noted
as soon as the first X-ray maps of clusters became available
\citep{Cowie1977, Fabian1977} and it was made more acute when X-ray
spectroscopy demonstrated that very little gas is cooling through
temperatures even a factor of three below that of the bulk of the gas
\citep{Peterson2001, Tamura2001}. A clue to the solution may come from
the observation \citep{Burns1981} that every cluster with a strong
cooling flow also contains a massive and active central radio galaxy.
\cite{Tabor1993} suggested that radio galaxies might regulate cooling
flows, and this idea has gained considerable recent support from X-ray
maps which show direct evidence for an interaction between radio lobes
and the intracluster gas \citep{Fabian2003, McNamara2000,
McNamara2005}. A number of authors have suggested ways in which the
radio source might replace the thermal energy lost to X-ray emission
\citep{Binney1995, Churazov2002, Bruggen2002, Ruszkowski2002,
Kaiser2003, Omma2004b}. We do not focus on this aspect of the problem
here, but rather demonstrate that if the scaling properties of radio
source feedback are set so they can plausibly solve the cooling flow
problem they induce a cut-off at the bright end of the galaxy
luminosity function which agrees well with observation.

Another puzzling aspect of the galaxy population is the fact that the
most massive galaxies, typically ellipticals in clusters, are made of
the oldest stars and so finished their star formation earlier than
lower mass galaxies \citep{Bender1999}. Confirming evidence for this
comes from look-back studies which show that both star-formation and
AGN activity take place more vigorously and in higher mass objects at
redshifts of 1 to 2 than in the present Universe
\citep[e.g.][]{Shaver1996, Madau1996}.  \cite{Cowie1996} termed this
phenomenon `down-sizing', and {\it prima facie} it conflicts with
hierarchical growth of structure in a $\Lambda$CDM cosmogony where
massive dark haloes assemble at lower redshift than lower mass haloes
\citep[e.g.][]{Lacey1993}. This puzzle is related to the previous two;
the late-forming high mass haloes in $\Lambda$CDM correspond to groups
and clusters of galaxies, and simple theories predict that at late
times their central galaxies should grow to masses larger than those
observed through accretion from cooling flows. In the model we present
below, radio galaxies prevent significant accretion, thus limiting the
mass of the central galaxies and preventing them from forming stars at
late times when their mass and morphology can still change through
mergers. The result is a galaxy luminosity function with a sharper
high-mass cut-off in which the most massive systems are red, dead and
elliptical.

To make quantitative predictions for the galaxy population in a
$\Lambda$CDM universe it is necessary to carry out simulations.
Present numerical capabilities allow reliable simulation of the
coupled nonlinear evolution of dark matter and diffuse gas, at least
on the scales which determine the global properties of galaxies.  Once
gas cools and condenses into halo cores, however, both its structure
and the rates at which it turns into stars and feeds black holes are
determined by small-scale `interstellar medium' processes which are
not resolved. These are usually treated through semi-analytic recipes,
parameterised formulae which encapsulate `subgrid' physics in terms of
star formation thresholds, Schmidt `laws' for star formation, Bondi
models for black hole feeding, etc. The form and the parameters of
these recipes are chosen to reproduce the observed systematics of star
formation and AGN activity in galaxies \citep[e.g.][]{Kennicutt1998}.
With a well-constructed scheme it is possible to produce stable and
numerically converged simulations which mimic real star-forming
galaxies remarkably well \citep{Springel2003b}. In strongly
star-forming galaxies, massive stars and supernovae produce winds
which redistribute energy, mass and heavy elements over large regions
\citep{Heckman1990, Martin1999}. Even stronger feedback is possible,
in principle, from AGN \citep{Begelman1991}. In both cases the
determining processes occur on very small scales and so have to be
included in simulations through parametrised semi-analytic
models. Unfortunately, the properties of simulated galaxies turn out
to depend strongly on how these unresolved star-formation and feedback
processes are treated.

Since the diffuse gas distribution and its cooling onto galaxies are
strongly affected by the description adopted for the subgrid physics,
every modification of a semi-analytic model (or of its parameters)
requires a simulation to be repeated. This makes parameter studies or
tests of, say, the effects of different AGN feedback models into a
very expensive computing exercise. A cost-effective alternative is to
represent the behaviour of the diffuse gas also by a semi-analytic
recipe. Since the dark matter couples to the baryons only through
gravity, its distribution on scales of galaxy haloes and above is only
weakly affected by the details of galaxy formation. Its evolution can
therefore be simulated once, and the evolution of the baryonic
component can be included in post-processing by applying semi-analytic
models to the stored histories of all dark matter objects.  Since the
second step is computationally cheap, available resources can be used
to carry out the best possible dark matter simulation, and then many
parameter studies or tests of alternative models can be carried out in
post-processing. This simulation approach was first introduced by
\cite{Kauffmann1999} and it is the approach we apply here to the
Millennium Run, the largest calculation to date of the evolution of
structure in the concordance $\Lambda$CDM cosmogony
\citep{Springel2005}.

This paper is organised as follows.  In Section~\ref{darkmatter} we
describe the Millennium Run and the post-processing we carried out to
construct merging history trees for all the dark haloes within
it. Section~\ref{sam} presents the model for the formation and
evolution of galaxies and their central supermassive black holes that
we implement on these merging trees.  Section \ref{results} describes
the main results of our modelling, concentrating on the influence of
`radio mode' feedback on the properties of the massive galaxy
population.  In Section \ref{discussion} we discuss physical models
for black hole accretion which may explain the phenomenology required
for our model to be successful.  Finally, Section \ref{conclusion}
summarises our conclusions and suggests some possible directions for
future investigation.

\section{The dark matter skeleton: the Millennium Run}
\label{darkmatter}

\begin{figure*}
\centerline{\psfig{file=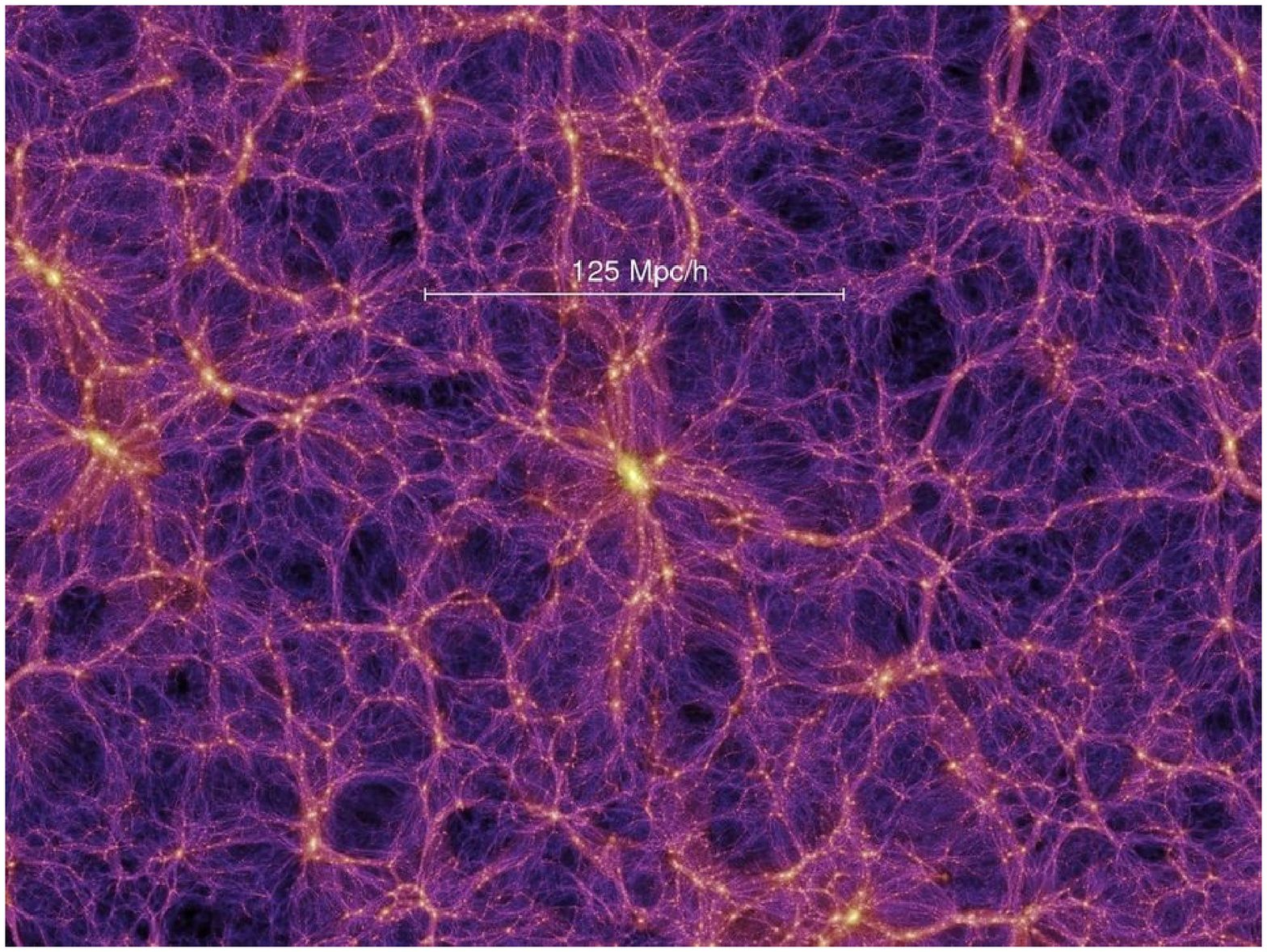,width=0.83\textwidth}}
\centerline{\psfig{file=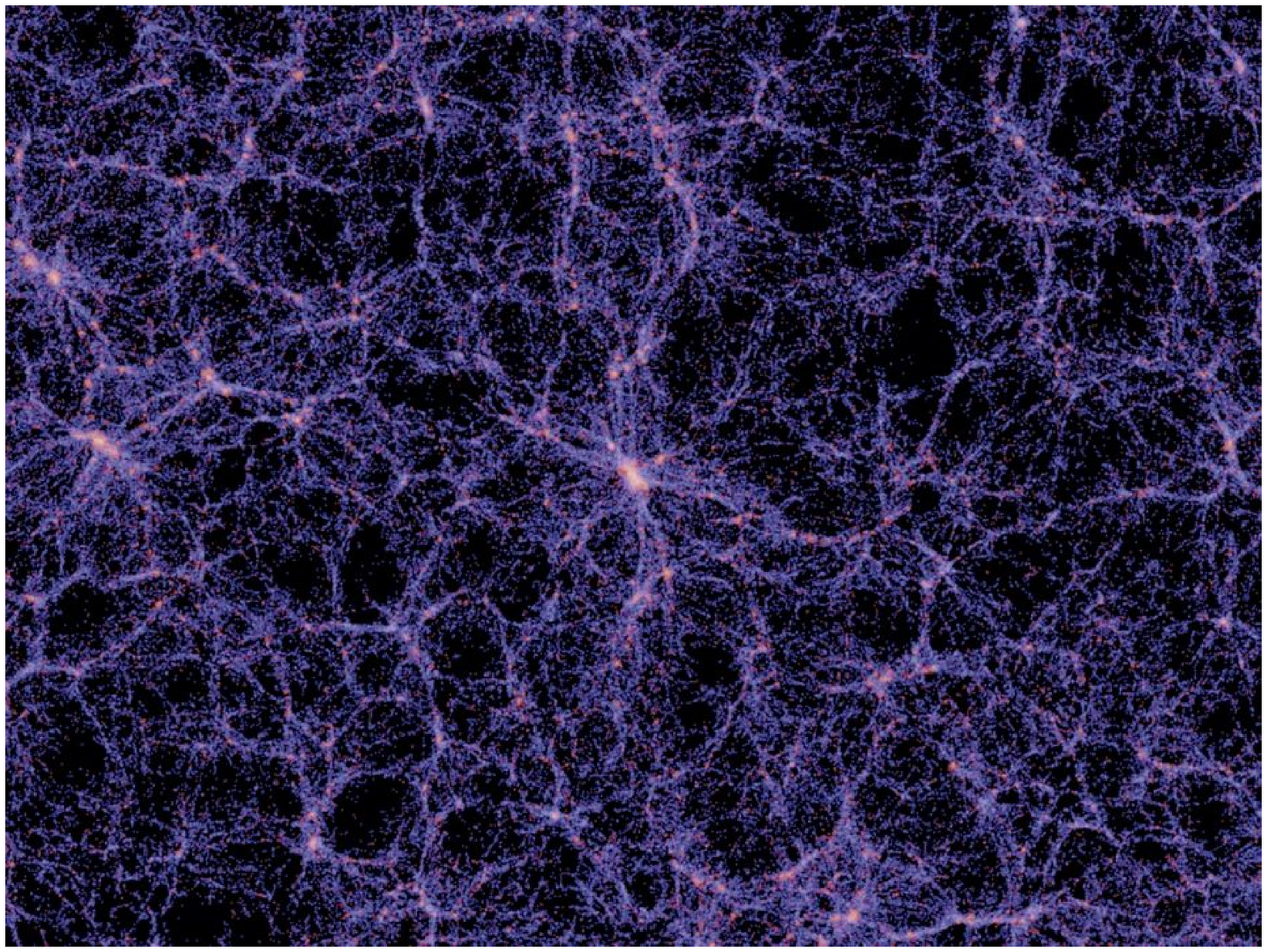,width=0.83\textwidth}}
\caption{The redshift zero distribution of dark matter (top) and of galaxy
 light (bottom) for a slice of thickness $15\,h^{-1}$Mpc, cut from the
 Millennium Run. For the dark matter distribution, intensity encodes surface
 density and colour encodes local velocity dispersion. For the light
 distribution, intensity encodes surface brightness and colour encodes mean
 B$-$V colour. The linear scale is shown by the bar in the top panel.}

\label{structure}
\end{figure*}

Our model for the formation and evolution of galaxies and their
central supermassive black holes is implemented on top of the
Millennium Run, a very large dark matter simulation of the concordance
$\Lambda$CDM cosmology with $2160^3\simeq 1.0078\times 10^{10}$
particles in a periodic box of $500\,h^{-1}$Mpc on a side. A full
description is given in \citet{Springel2005}; here we summarise the
main simulation characteristics and the definition and construction of
the dark matter merging history trees we use in our galaxy formation
modelling.  The dark matter distribution is illustrated in the top
panel of Fig.~\ref{structure} for a $330\times 280\times
15\,h^{-1}$Mpc slice cut from the full volume.  The projection is
colour coded by density and local velocity dispersion, and illustrates
the richness of dark matter structure for comparison with structure in
the light distribution to which we will come later. Dark matter plots
on a wider range of scales may be found in \citet{Springel2005}.

\subsection{Simulation characteristics}

We adopt cosmological parameter values consistent with a combined
analysis of the 2dFGRS \citep{Colless2001} and first-year WMAP data
\citep{Spergel2003,Seljak2005}. They are $\Omega_{\rm m}= \Omega_{\rm
dm}+\Omega_{\rm b}=0.25$, $\Omega_{\rm b}=0.045$, $h=0.73$,
$\Omega_\Lambda=0.75$, $n=1$, and $\sigma_8=0.9$. Here $\Omega_{\rm
m}$ denotes the total matter density in units of the critical density
for closure, $\rho_{\rm crit}=3 H_0^2/(8\pi G)$. Similarly,
$\Omega_{\rm b}$ and $\Omega_\Lambda$ denote the densities of baryons
and dark energy at the present day. The Hubble constant is given as
$H_0 = 100\, h\, {\rm km\, s^{-1} Mpc^{-1}}$, while $\sigma_8$ is the
{\em rms} linear mass fluctuation within a sphere of radius $8\,
h^{-1}{\rm Mpc}$ extrapolated to $z\!=\!0$.

The chosen simulation volume is a periodic box of size
$500\,h^{-1}{\rm Mpc}$, which implies a particle mass of $8.6\times
10^8\,h^{-1}{\rm M}_{\odot}$.  This volume is large enough to include
interesting objects of low space density, such as quasars or rich
galaxy clusters, the largest of which contain about 3 million
simulation particles at $z\!=\!0$. At the same time, the mass
resolution is sufficient that haloes that host galaxies as faint as
$0.1\,L_\star$ are typically resolved with at least $\sim\!100$
particles.  Note that although discreteness noise significantly
affects the merger histories of such low mass objects, the galaxies
that reside in halos with $\simlt\!100$ particles are usually
sufficiently far down the luminosity function that any uncertainty in
their properties has little impact on our results or conclusions.

The initial conditions at $z\!=\!127$ were created by displacing particles from a
homogeneous, `glass-like' distribution \citep{White1996} using a Gaussian
random field with a $\Lambda$CDM linear power spectrum as given by the
Boltzmann code {\small CMBFAST} \citep{Seljak1996}.  The simulation was then
evolved to the present epoch using a leapfrog integration scheme with
individual and adaptive time steps, with up to $11\,000$ time steps for
individual particles.

The simulation itself was carried out with a special version of the {\small
GADGET-2} code \citep{Springel2001b, Springel2005Gadget2} optimised for very
low memory consumption so that it would fit into the nearly 1~TB of physically
distributed memory available on the parallel IBM p690 computer\footnote{This
computer is operated by the Computing Centre of the Max-Planck Society in
Garching, Germany.} used for the calculation.  The computational algorithm
used the `TreePM' method \citep{Xu1995,Bode2000,Bagla2002} to evaluate
gravitational forces, combining a hierarchical multipole expansion, or `tree'
algorithm \citep{Barnes1986}, and a classical, Fourier transform particle-mesh
method \citep{Hockney1981}.  An explicit force-split in Fourier-space produces
a very nearly isotropic force law with negligible force errors at the force
matching scale.  The short-range gravitational force law is softened on
comoving scale $5\,h^{-1}{\rm kpc}$ (Plummer-equivalent) which may be taken as
the spatial resolution limit of the calculation, thus achieving a dynamic
range of $10^5$ in 3D. The calculation, performed in parallel on 512
processors, required slightly less than $350\,000$ processor hours of CPU
time, or 28 days of wall-clock time.

\subsection{Haloes, substructure, and merger tree construction}

Our primary application of the Millennium Run in this paper uses
finely resolved hierarchical merging trees which encode the full
formation history of tens of millions of haloes and the subhaloes that
survive within them. These merging history trees are the backbone of
the model that we implement in post-processing in order to simulate
the wide range of baryonic processes that are important during the
formation and evolution of galaxies and their central supermassive
black holes.

We store the full particle data between $z\!=\!20$ and $z\!=\!0$ at 60
output times spaced in expansion factor according to the formula 
\begin{equation}
\log \, (1+z_n) =  n \, (n+35)/4200~.
\end{equation}
This spacing is `locally' logarithmic but becomes smoothly finer at
lower redshift, with a temporal resolution by redshift zero of
approximately $300\,$Myr.  Additional outputs are added at
$z\!=\!30,50,80,127$ to produce a total of 64 snapshots in all. We
note that each snapshot has a total size in excess of 300 GB, giving a
raw data volume of nearly 20 TB.

Together with each particle coordinate dump, the simulation code directly
produces a friends-of-friends (FOF) group catalogue on the fly and in
parallel. FOF groups are defined as equivalence classes where any pair of two
particles is placed into the same group if their mutual separation is less
than 0.2 of the mean particle separation \citep{Davis1985}. This criterion
combines particles into groups with a mean overdensity of about 200,
corresponding approximately to that expected for a virialised group.  The
group catalogue saved to disk for each output only kept groups with at least
20 particles.

High-resolution simulations like the present one exhibit a rich
substructure of gravitationally bound dark matter subhaloes orbiting
within larger virialised halos \citep[e.g.][]{Ghigna1998}. The FOF
group-finder built into our simulation code identifies the haloes but
not their subhalos. Since we wish to follow the fate of infalling
galaxies and their halos, and these are typically identifiable for a
substantial time as a dark matter subhalo within a FOF halo, we apply
in post-processing an improved and extended version of the {\small
SUBFIND} algorithm of \citet{Springel2001}. This computes an
adaptively smoothed dark matter density field within each halo using a
kernel-interpolation technique, and then exploits the topological
connectivity of excursion sets above a density threshold to identify
substructure candidates. Each substructure candidate is subjected to a
gravitational unbinding procedure. If the remaining bound part has
more than 20 particles, the subhalo is kept for further analysis and
some of its basic physical properties are determined (angular
momentum, maximum of its rotation curve, velocity dispersion,
etc.). After all subhaloes are identified they are extracted from the
FOF halo so that the remaining featureless `background' halo can also
be subjected to the unbinding procedure. This technique, however,
neglects the fact that substructures embedded within a halo help to
bind its material, and thus removing them can, in principal, unbind
some of the FOF halo that may otherwise still be loosely bound.  We
accept this small effect for technical reasons related to the
robustness of our halo definition procedures and the need for
unambigous particle-subhalo assignments in our data structures. We
note that the total mass in substructures is typically below 10\% and
often substantially smaller.  Thus any bias in the bound mass of the
parent halo due to additional unbinding is always very small.

To compute a virial mass estimate for each FOF halo we use the
spherical-overdensity approach, where the centre is determined using
the minimum of the gravitational potential within the group and we
define the boundary at the radius which encloses a mean overdensity of
200 times the critical value. The virial mass, radius and circular
velocity of a halo at redshift $z$ are then related by
\begin{equation}
M_{\rm vir}\ =\ \frac{100}{\rm G}\, H^2(z)\, R_{\rm vir}^3\ =\
\frac{V_{\rm vir}^3}{10\,{\rm G}\,H(z)}
\label{virial}
\end{equation}
where $H(z)$ is the Hubble constant at redshift $z$.

At $z\!=\!0$ our procedures identify $17.7\times 10^6$ FOF groups,
down from a maximum of $19.8\times 10^6$ at $z\!=\!1.4$ when groups
were more abundant but of lower mass on average.  The $z\!=\!0$ groups
contain a total of $18.2\times 10^6$ subhaloes, with the largest FOF
group containing 2328 of them. (Note that with our definitions, all
FOF groups contain at least one subhalo, the {\it main} subhalo which
is left over after removal of any substructure and the unbound
component. We require this main subhalo to contain at least 20
particles.)

Having found all haloes and subhaloes at all output snapshots, we then
characterise structural evolution by building merging trees that describe in
detail how haloes grow as the universe evolves.  Because structures merge
hierarchically in CDM universes there can be several progenitors for any given
halo, but in general there is only one descendant. (Typically the cores of
virialised dark matter structures do not split into two or more objects.) We
therefore construct merger trees based on defining a unique descendant for
each halo and subhalo.  This is, in fact, sufficient to define the entire
merger tree, since the progenitor information then follows implicitly.
Further details can be found in \citet{Springel2005}.

We store the resulting merging histories tree by tree. Since each tree
contains the full formation history of some $z\!=\!0$ halo, the
physical model for galaxy formation can be computed sequentially tree
by tree.  It is thus unnecessary to load all the history information
on the simulation into computer memory at the same time. Actually,
this would be currently impossible, since the combined trees contain a
total of around 800 million subhaloes for each of which a number of
attributes are stored. Note that, although evolving the galaxy
population sequentially in this way allows us to consider, in
principle, all interactions between galaxies that end up in the same
present-day FOF halo, it does not allow us to model longer range
interactions which might take place between galaxies that end up in
different FOF halos at $z\!=\!0$.

\section{Building galaxies: the semi-analytic model}
\label{sam}

\begin{table*}
\centering
\caption{A summary of our main model parameters and their best values and
  plausible ranges, as described in the text.  Once set, these values are kept
  fixed for all results presented in this paper, in particular for models in
  which AGN feedback is switched off.}
\begin{tabular}{clcc}
\hline \hline
parameter               & description                                                               & best value  & plausible range    \\
\hline \hline
$f_b$                   & cosmic baryon fraction ($\S\ref{reionization}$)                           & $0.17$      & fixed              \\
$z_0$, $z_r$            & redshift of reionization ($\S\ref{reionization}$)                         & $8$, $7$    & fixed              \\
\hline
$f_{\rm{BH}}$           & merger cold gas BH accretion fraction ($\S\ref{quasar}$)                  & $0.03$      & $0.02 - 0.04$      \\
$\kappa_{\rm{AGN}}$     & quiescent hot gas BH accretion rate (${\rm M_{\odot} yr^{-1}}$) ($\S\ref{radio}$) & $6\times 10^{-6}$ & $(4 - 8) \times 10^{-6}$ \\
\hline
$\alpha_{\rm{SF}}$      & star formation efficiency ($\S\ref{SF}$)                                  & $0.07$      & $0.05 - 0.15$      \\
\hline
$\epsilon_{\rm{disk}}$  & SN feedback disk reheating efficiency ($\S\ref{SN}$)                      & $3.5$       & $1 - 5$            \\
$\epsilon_{\rm{halo}}$  & SN feedback halo ejection efficiency ($\S\ref{SN}$)                       & $0.35$      & $0.1 - 0.5$        \\
$\gamma_{\rm{ej}}$      & ejected gas reincorporation efficiency ($\S\ref{SN}$)                     & $0.5$       & $0.1 - 1.0$        \\
\hline
$T_{\rm{merger}}$       & major merger mass ratio threshold ($\S\ref{mergers}$)                     & $0.3$       & $0.2 - 0.4$        \\
\hline
$R$                     & instantaneous recycled fraction of SF to the cold disk ($\S\ref{metals}$) & $0.3$       & $0.2 - 0.4$        \\
$Y$                     & yield of metals produced per unit SF ($\S\ref{metals}$)                   & $0.03$      & $0.02 - 0.04$      \\
\hline \hline
\end{tabular}
\label{parameters}
\end{table*}

\subsection{Overview}
\label{overview}

In the following sub-sections we describe the baryonic physics of one
particular model for the formation and evolution of galaxies and of their
central supermassive black holes.  A major advantage of our simulation
strategy is that the effects of parameter variations within this model (or
indeed alternative assumptions for some of the processes) can be explored at
relatively little computational expense since the model operates on the stored
database of merger trees; the simulation itself and the earlier stages of
post-processing do not need to be repeated. We have, in fact, explored a wide
model and parameter space to identify our current best model.  We summarise
the main parameters of this model, their `best' values, and their plausible
ranges in Table~\ref{parameters}.  These choices produce a galaxy population
which matches quite closely many observed quantities. In this paper we discuss
the field galaxy luminosity-colour distribution; the mean stellar mass --
stellar age relation; the Tully-Fisher relation, cold gas fractions and
gas-phase metallicities of Sb/c spirals; the colour -- magnitude relation of
ellipticals; the bulge mass -- black hole mass relation; and the
volume-averaged cosmic star-formation and black hole accretion histories. In
\citet{Springel2005} we also presented results for galaxy correlations as a
function of absolute magnitude and colour, for the baryonic `wiggles' in the
large-scale power spectrum of galaxies, and for the abundance, origin and fate
of high redshift supermassive black holes which might correspond to the $z\sim
6$ quasars discovered by the SDSS \citep{Fan2001}

In our model we aim to motivate each aspect of the physics of galaxy
formation using the best available observations and simulations.  Our
parameters have been chosen to reproduce local galaxy properties and
are stable in the sense that none of our results is critically
dependent on any single parameter choice; plausible changes in one
parameter or recipe can usually be accommodated through adjustment of
the remaining parameters within their own plausible range. The
particular model we present is thus not unique. Importantly, our model
for radio galaxy heating in cooling flows, which is the main focus of
this paper, is only weakly influenced by the remaining galaxy
formation and black hole growth physics.  This is because our radio
mode feedback is active only in massive objects and at late times and
it has no effect during the principal growth phase of most galaxies
and AGN.

The distribution of galaxy light in our `best' model is shown in the bottom
panel of Fig.~\ref{structure} for comparison with the mass distribution in the
top panel.  For both the volume is a projected $330\times 280\times
15\,h^{-1}$Mpc slice cut from the full $0.125\,h^{-3}$Gpc$^3$ simulation
box. The plot of surface brightness is colour-coded by the luminosity-weighted
mean ${\rm B\!-\!V}$ colour of the galaxies.  On large scales light clearly
follows mass, but non-trivial biases become evident on smaller scales,
especially in `void' regions. The redder colour of galaxies in high density
regions is also very clear.

\subsection{Gas infall and cooling}
\label{cooling}

We follow the standard paradigm set out by \citet{White1991} as
adapted for implementation on high resolution N-body simulations by
\citet{Springel2001} and \citet{deLucia2004}. This assumes that as
each dark matter halo collapses, its own `fair share' of cosmic
baryons collapse with it (but see Section~\ref{reionization} below).
Thus in our model the mass fraction in baryons associated with every
halo is taken to be $f_b=17\%$, consistent with the first-year WMAP
result \citep{Spergel2003}. Initially these baryons are in the form of
diffuse gas with primordial composition, but later they include gas in
several phases as well as stars and heavy elements. The fate of the
infalling gas depends on redshift and on the depth of the halo
potential \citep{Silk1977, Rees1977, Binney1977, White1978}.  At late
times and in massive systems the gas shocks to the virial temperature
and is added to a quasi-static hot atmosphere that extends
approximately to the virial radius of the dark halo. Gas from the
central region of this atmosphere may accrete onto a central object
through a cooling flow. At early times and in lower mass systems the
infalling gas still shocks to the virial temperature but its
post-shock cooling time is sufficiently short that a quasi-static
atmosphere cannot form. Rather the shock occurs at much smaller radius
and the shocked gas cools rapidly and settles onto a central object,
which we assume to be a cold gas disk. This may in turn be subject to
gravitational instability, leading to episodes of star formation.

More specifically, the cooling time of a gas is conventionally taken as the
ratio of its specific thermal energy to the cooling rate per unit volume,
\begin{equation}
t_{\rm cool} = \frac{3}{2} \frac{\bar{\mu} m_p k T }{ \rho_g (r) \Lambda
(T,Z)} ~.
\label{tcool}
\end{equation}
Here $\bar{\mu} m_p$ is the mean particle mass, $k$ is the Boltzmann constant,
$\rho_g(r)$ is the hot gas density, and $\Lambda (T,Z)$ is the cooling
function.  The latter depends both on the metallicity $Z$ and the temperature
of the gas. In our models we assume the post-shock temperature of the
infalling gas to be the virial temperature of the halo, $T=35.9\,(V_{\rm
vir}/\rm{km\,s^{-1}})^2$K.  When needed, we assume that the hot gas within a
static atmosphere has a simple `isothermal' distribution,
\begin{equation}
\rho_g(r) = \frac{m_{\rm hot}}{4 \pi R_{\rm vir} r^2} ~,
\label{rhog}
\end{equation}
where $m_{\rm hot}$ is the total hot gas mass associated with the halo
and is assumed to extend to its virial radius $R_{\rm vir}$.  

\begin{figure*}
\plotfull{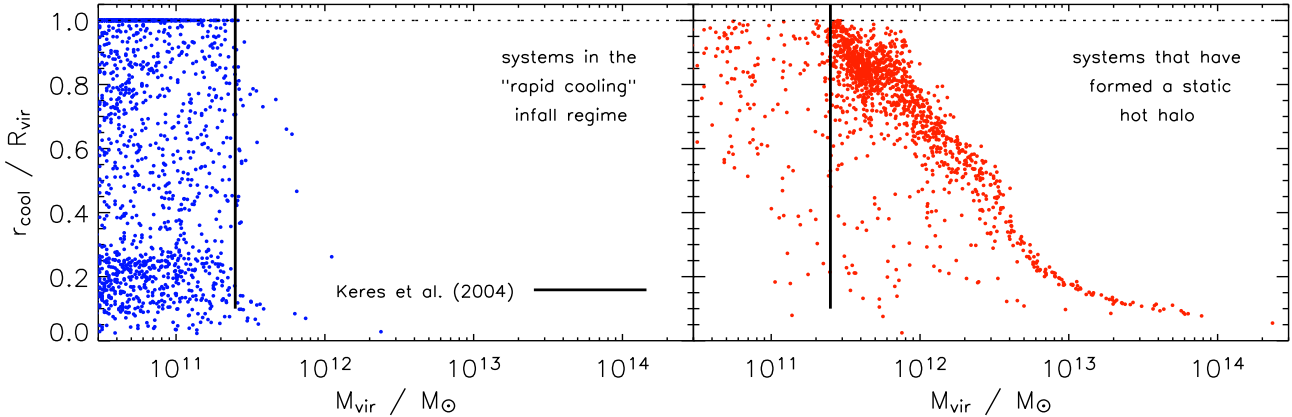}
\caption{The ratio of the cooling radius to virial radius for a random
  selection of virialised systems at $z\!=\!0$ plotted against their
  dark matter virial mass.  Systems identified to be in the `rapid
  cooling' regime are shown in the left panel, while those that have
  formed a static hot halo are shown on the right
  (Section~\ref{cooling}).  A sharp transition between the two regimes
  is seen close to that found by \citet{Keres2004}, marked by the
  solid vertical line.  }
\label{rcool}
\end{figure*}

To estimate an instantaneous cooling rate onto the central object of a
halo, given its current hot gas content, we define the cooling radius,
$r_{\rm cool}$, as the radius at which the local cooling time
(assuming the structure of Eq.~\ref{rhog}) is equal to a suitably
defined age for the halo. \cite{White1991} took this age to be the
Hubble time $t_{\rm H}$, while \cite{Cole1994, Cole2000} used the time
scale over which the main progenitor last doubled its mass.
\cite{Somerville1999} argued that the time interval since the last
major merger is more appropriate since such mergers redistribute hot
gas within the halo.  Here we follow \cite{Springel2001} and
\cite{deLucia2004} and define the cooling radius as the point where
the local cooling time is equal to the halo dynamical time, $R_{\rm
vir}/V_{\rm vir}= 0.1\, H(z)^{-1}$. This is an order of magnitude
smaller than $t_{\rm H}$ and so results in substantially smaller
cooling radii and cooling rates (typically by a factor of 3) than the
assumption of \cite{White1991}. Our choice is justified by the tests
of \cite{Yoshida2002} who verified explicitly that it results in good
object-by-object agreement between the amount of gas predicted to
condense in galaxy mass haloes and the amount which actually condensed
in their high-resolution SPH simulations of the formation of a galaxy
cluster and its environment. These tests assumed primordial abundances
in the cooling function. When we implement a chemical enrichment model
consistent with the observed element abundances in intracluster gas
(see Section~\ref{metals}) cooling rates in galaxy mass haloes are
substantially enhanced and we find (as did \citealt{deLucia2004}) that
a smaller coefficient than used in the original \citeauthor{White1991}
cooling model is required to avoid excessive condensation of gas.

Using the above definition, a cooling rate can now be determined
through a simple continuity equation,
\begin{equation}
\dot{m}_{\rm cool} = 4 \pi \rho_g(r_{\rm cool}) r_{\rm cool}^2
\dot{r}_{\rm cool} ~. 
\label{mdotcool}
\end{equation}
Despite its simplicity, this equation is a good approximation to
the rate at which gas is deposited at the centre in the
\citet{Bertschinger1989} similarity solution for a cooling flow. Putting it
all together we take the cooling rate within a halo containing a hot
gas atmosphere to be
\begin{equation}
\dot{m}_{\rm cool} = 0.5\, m_{\rm hot}\frac{r_{\rm
    cool}{V_{\rm vir}}}{R_{\rm vir}^2} ~. 
\label{coolstatic}
\end{equation}
We assume this equation to be valid when $r_{\rm cool} < R_{\rm vir}$.
This is the criterion which \citet{White1991} proposed to separate
the \emph{static hot halo regime} from the \emph{rapid cooling regime}
(see below). 

In low mass haloes or at high redshifts the formal cooling radius lies
outside the virial radius $r_{\rm cool} > R_{\rm vir}$.  The
post-shock gas then cools in less than one sound crossing time and
cannot maintain the pressure needed to support an accretion shock at
large radius. The high-resolution spherical infall simulations of
\cite{ForcadaMiro1997} show that in this situation the accretion shock
moves inwards, the post-shock temperature {\it increases} and the mass
stored in the post-shock hot atmosphere {\it decreases}, because the
post-shock gas rapidly cools onto the central object.  In effect, all
infalling material is accreted immediately onto the central disk. In
this \emph{rapid cooling regime} we therefore set the cooling rate
onto the central object to be equal to the rate at which new diffuse
gas is added to the halo.

\subsubsection{Rapid cooling or cold accretion?}
\label{coldaccretion}
Although much simplified, the above model was shown by
\cite{Yoshida2002} and \cite{Helly2003} to give reasonably accurate,
object-by-object predictions for the cooling and accumulation of gas
within the galaxies that formed in their $N$-body+SPH
simulations. These neglected star-formation and feedback effects in
order to test the cooling model alone.  They also 
assumed primordial abundances in the cooling function.  In
Fig.~\ref{rcool} we show the ratio $r_{\rm cool}/R_{\rm vir}$ as a
function of virial mass for haloes in the `rapid cooling' regime (left
panel) and in the `static halo' regime (right panel) at $z\!=\!0$ for
our `best' galaxy formation model.  The two regimes are distinguished
by the dominant gas component in each halo: when the mass of hot halo
gas exceeds that of cold disk gas, we say the galaxy has formed a
static halo, otherwise the system is taken to be in the rapid cooling
phase.  Many haloes classified as `rapidly cooling' by this criterion
have $r_{\rm cool} < R_{\rm vir}$, which would apparently indicate a
static hot halo.  This is misleading, however, as systems where
cooling is rapid deposit infalling gas onto the central galactic disk
on a short timescale until they have a low-mass residual halo which
satisfies $r_{\rm cool}\sim R_{\rm vir}$. This then persists until the
next infall event. Averaging over several cycles of this behaviour,
one finds that the bulk of the infalling gas cools rapidly.  This is
why we choose to classify systems by their dominant gas component.
Note also that a massive hot halo forms immediately once cooling
becomes inefficient, just as in the 1-D infall simulations of
\cite{ForcadaMiro1997} and \cite{Birnboim2003}. Our classification is
thus quite robust.

The transition between the `rapid cooling' and `static halo' regimes
is remarkably well defined. At $z\!=\!0$ it occurs at a halo virial
mass of $2$--$3\times 10^{11} M_{\odot}$, and is approximately
independent of redshift out to at least $z\!=\!6$.  This is close to
the transition mass found for the same cosmology by
\cite{Birnboim2003} using spherically symmetric simulations, and by
\cite{Keres2004} using fully 3-D simulations.  This is, in fact, a
coincidence since both sets of authors assume cooling functions with
no metals, whereas our `best' model includes heavy elements which
substantially enhance cooling at the temperatures relevant for the
transition.  Enrichment in the `rapid cooling' regime arises from the
mixing of infalling primordial gas with supernova-enriched gas ejected
in a galactic wind (see Section~\ref{SN}).  If we modify our model to
assume a zero-metal cooling function, we find the transition mass to
shift downwards by about a factor of 2--3, resulting in a lower
cooling rate in comparison to the work of \cite{Birnboim2003} and
\cite{Keres2004}.

The reason for the more efficient (we would argue overly efficient)
cooling appears to be different in these two studies. The spherical
infall simulations of \cite{ForcadaMiro1997} showed good agreement
with a transition at the point predicted using the original cooling
radius definition of \cite{White1991} rather than our revised
definition which was checked explicitly against SPH simulations by
\cite{Yoshida2002}.  Spherical models thus predict more efficient
cooling than occurs in typical 3-D situations.  This explains,
perhaps, why \cite{Birnboim2003} find a higher transition mass than we
would predict for zero metallicity. \cite{Yoshida2002} also showed
that the density estimation in the implementation of SPH used by
\cite{Keres2004} leads to overcooling of gas in galaxy mass objects as
compared to their own entropy and energy conserving SPH scheme; the
effect is large enough to explain why \cite{Keres2004} find a higher
transition mass than we find for their assumed cooling function.

Both \citeauthor{Birnboim2003} and \citeauthor{Keres2004} refer to the
`rapid cooling' regime as `cold infall'.  This is, in fact, a
misnomer. In this mode the accretion shock occurs closer to the
central object, and so deeper in its potential well than when there is
a static hot halo.  As a result, the pre-shock velocity of infalling
gas is greater, resulting in a {\it larger} post-shock
temperature. The two modes do not differ greatly in the temperature to
which infalling gas is shocked, but rather in how long (compared to
the system crossing time) the gas spends at the post-shock temperature
before its infall energy is lost to radiation.  Finally, we note that
the existence and importance of these two modes was the major insight
of the original work of \cite{Silk1977}, \cite{Binney1977} and
\cite{Rees1977} and has been built into all modern theories for galaxy
formation. A detailed discussion can be found, for example, in
\cite{White1991}.

\subsection{Reionization}
\label{reionization}

Accretion and cooling in low mass haloes is required to be inefficient to
explain why dwarf galaxies contain a relatively small fraction of all
condensed baryons \citep{White1978}.  This inefficiency may in part result
from photoionisation heating of the intergalactic medium (IGM) which
suppresses the concentration of baryons in shallow potentials
\citep{Efstathiou1992}.  More recent models identify the possible low-redshift
signature of such heating in the faint end of the galaxy luminosity function,
most notably in the abundance of the dwarf satellite galaxies in the local
group \citep[e.g.][]{Tully2002, Benson2002}.

\cite{Gnedin2000} showed that the effect of photoionization heating on the gas
content of a halo of mass $M_{\rm vir}$ can be modelled by defining a
characteristic mass scale, the so called \emph{filtering mass}, $M_F$, below
which the gas fraction $f_b$ is reduced relative to the universal value:
\begin{equation}
f_{\rm b}^{\rm halo}(z,M_{\rm vir}) = \frac{f_{\rm b}^{\rm cosmic}}{(1
  + 0.26 \,M_{\rm F}(z) / M_{\rm vir})^3}~.
\label{reion}
\end{equation}
The filtering mass is a function of redshift and changes most
significantly around the epoch of reionization. It was estimated by
Gnedin using high-resolution $\rm{SLH}$-$\rm{P^3M}$ simulations
\citep[but see][]{Hoeft2005}.  \cite{Kravtsov2004} provided an
analytic model for these results which distinguishes three `phases' in
the evolution of the IGM: $z\!>\!z_0$, the epoch before the first
$\rm{H II}$ regions overlap; $z_0\!<\!z\!<\!z_r$, the epoch when
multiple $\rm{H II}$ regions overlap; $z\!<\!z_r$, the epoch when the
medium is almost fully reionized.  They find that choosing $z_0\!=\!8$
and $z_r\!=\!7$ provides the best fit to the numerically determined
filtering mass.  We adopt these parameters and keep them fixed
throughout our paper.  This choice results in a filtering mass of $4
\times 10^{9} M_{\odot}$ at $z\!=\!z_r$, and $3 \times 10^{10}
M_{\odot}$ by the present day.  See Appendix~B of \cite{Kravtsov2004}
for a full derivation and description of the analytic model.

\subsection{Black hole growth, AGN outflows, and cooling suppression}
\label{blackhole}

There is a growing body of evidence that active galactic nuclei (AGN) are a
critical piece in the galaxy formation puzzle.  Our principal interest in this
paper is their role in suppressing cooling flows, thereby modifying the
luminosities, colours, stellar masses and clustering of the galaxies that
populate the bright end of the galaxy luminosity function.  To treat this
problem, we first need a physical model for the growth of black holes within
our galaxies.

\subsubsection{The `quasar mode'}
\label{quasar}

In our model (which is based closely on that of
\citealt{Kauffmann2000}) supermassive black holes grow during galaxy
mergers both by merging with each other and by accretion of cold disk
gas.  For simplicity, black hole coalescence is modelled as a direct
sum of the progenitor masses and thus ignores gravitational wave
losses (including such losses is found to have little effect on the
properties of the final galaxy population).  We assume that the gas
mass accreted during a merger is proportional to the total cold gas
mass present, but with an efficiency which is lower for smaller mass
systems and for unequal mergers.  Specifically,
\begin{equation}
\Delta m_{\rm BH,Q} = \frac{f'_{\rm BH} \ m_{\rm cold}}{1 +
  (280\,\rm{km\,s^{-1}}/V_{\rm vir})^2}~, 
\label{accretionQ}
\end{equation}
where we have changed the original parameterisation by taking
\begin{equation}
f'_{\rm BH} = f_{\rm BH}\ (m_{\rm sat}/m_{\rm central})~.
\label{fBHparameter}
\end{equation}
Here $f_{\rm BH}\approx 0.03$ is a constant and is chosen to reproduce the
observed local $m_{\rm BH}-m_{\rm bulge}$ relation \citep{Magorrian1998,
Marconi2003, Haring2004}.  In contrast to \cite{Kauffmann2000} we allow black
hole accretion during both major \emph{and} minor mergers although the
efficiency in the latter is lower because of the $m_{\rm sat}/m_{\rm central}$
term.  Thus, any merger-induced perturbation to the gas disk (which might come
from a bar instability or a merger-induced starburst -- see Section
\ref{mergers}) can drive gas onto the central black hole.  In this way, minor
merger growth of the black hole parallels minor merger growth of the
bulge. The fractional contribution of minor mergers to both is typically 
quite small, so that accretion driven by major mergers is the dominant 
mode of black hole growth in our model. We refer to this as the
`quasar mode'.  [Note that a more schematic treatment of black hole growth 
would suffice for the purposes of this paper, but in \cite{Springel2005} 
and in future work we wish to examine the build-up of the black hole population
within galaxies in considerably more detail.]

There is substantial evidence for strong hydrodynamic and radiative
feedback from optical/UV and X-ray AGN \citep{Arav2001, deKool2001,
Reeves2003, Crenshaw2003}. We have not yet explicitly incorporated
such feedback in our modelling, and it may well turn out to be
important (see, for example, the recent simulations of
\citealt{dimatteo2005, Springel2005c,Hopkins2005}). We assume `quasar mode' accretion to be
closely associated with starbursts, so this feedback channel may be
partially represented in our models by an enhanced effective feedback
efficiency associated with star formation and supernovae (see
Section~\ref{SN}).

\subsubsection{The `radio mode'}
\label{radio}
In our model, low energy `radio' activity is the result of hot gas accretion
onto a central supermassive black hole once a static hot halo has formed
around the black hole's host galaxy.  We assume this accretion to be continual
and quiescent and to be described by a simple phenomenological model:
\begin{equation}
\dot{m}_{\rm BH,R} = \kappa_{\rm{AGN}} \Big(\frac{m_{\rm BH}}{10^{8}
  M_{\odot}}\Big) \Big(\frac{f_{\rm hot}}{0.1}\Big)
  \Big(\frac{V_{\rm vir}}{200\,\rm{km\,s^{-1}}}\Big)^3 ~, 
\label{accretionR}
\end{equation}
where $m_{\rm BH}$ is the black hole mass, $f_{\rm hot}$ is the fraction of
the total halo mass in the form of hot gas, $V_{\rm vir} \propto T_{\rm
vir}^{1/2}$ is the virial velocity of the halo, and $\kappa_{\rm{AGN}}$ is a
free parameter with units of $\rm{M_{\odot} yr^{-1}}$ with which we control
the efficiency of accretion.  We find below that $\kappa_{\rm{AGN}} = 6\times
10^{-6} \rm{M_{\odot} yr^{-1}}$ accurately reproduces the turnover at the
bright end of the galaxy luminosity function.  Note that $f_{\rm hot}V_{\rm
vir}^3t_{\rm{H}}$ is proportional to the total mass of hot gas, so that our
formula is simply the product of the hot gas and black hole masses multiplied
by a constant efficiency and divided by the Hubble time $t_{\rm{H}}$. In fact, we
find $f_{\rm hot}$ to be approximately constant for $V_{\rm vir} \simgt
150\,{\rm km\,s^{-1}}$, so the dependence of $\dot{m}_{\rm BH,R}$ on this
quantity has little effect. The accretion rate given by Eq.~\ref{accretionR}
is typically orders-of-magnitude below the Eddington limit.  In
Section~\ref{discussion} we discuss physical accretion models which may
reproduce this phenomenology.

We assume that `radio mode' feedback injects sufficient energy into the 
surrounding medium to reduce or even stop the cooling flow described in
Section~\ref{cooling}.  We take the mechanical heating generated by
the black hole accretion of Eq.~\ref{accretionR} to be
\begin{equation}
L_{\rm BH} = \eta \,\dot{m}_{\rm BH} \,c^2 ~,
\label{energyR}
\end{equation}
where $\eta = 0.1$ is the standard efficiency with which mass is assumed
to produce energy near the event horizon, and $c$ is the speed of 
light.  This injection of energy compensates in part
for the cooling, giving rise to a modified infall rate
(Eq.~\ref{coolstatic}) of
\begin{equation}
\dot{m}_{\rm cool}' = \dot{m}_{\rm cool} - \frac{L_{\rm BH}}{\frac{1}{2}
V_{\rm vir}^2} ~.
\label{effective_cool}
\end{equation}
For consistency we never allow $\dot{m}_{\rm cool}'$ to fall below zero.  It
is worth noting that $\dot{m}_{\rm cool} \propto f_{\rm hot}^{3/2}\,
\Lambda(V_{\rm vir})^{1/2}\, V_{\rm vir}^2\, t_{\rm{H}}^{-1/2}$
(Eq.~\ref{coolstatic}) and $\dot{m}_{\rm heat} \equiv 2 L_{\rm BH}/V_{\rm vir}^2
\propto m_{\rm BH}\, f_{\rm hot}\, V_{\rm vir}$
(Eq.~\ref{effective_cool}), so that
\begin{equation}
\frac{\dot{m}_{\rm heat}}{\dot{m}_{\rm cool}} \propto \frac{m_{\rm BH}\,
t_{\rm{H}}^{1/2}}{f_{\rm hot}^{1/2}\,\Lambda(V_{\rm vir})^{1/2}\, 
V_{\rm vir}}\, .
\label{z_behaviour}
\end{equation}
(These scalings are exact for an EdS universe; we have omitted weak
coefficient variations in other cosmologies.)  Thus in our model the
effectiveness of radio AGN in suppressing cooling flows is greatest at late
times and for large values of black hole mass. This turns out to be the
qualitative behaviour needed for the suppression of cooling flows to reproduce
successfully the luminosities, colours and clustering of low redshift bright
galaxies.

\begin{figure}
\plotone{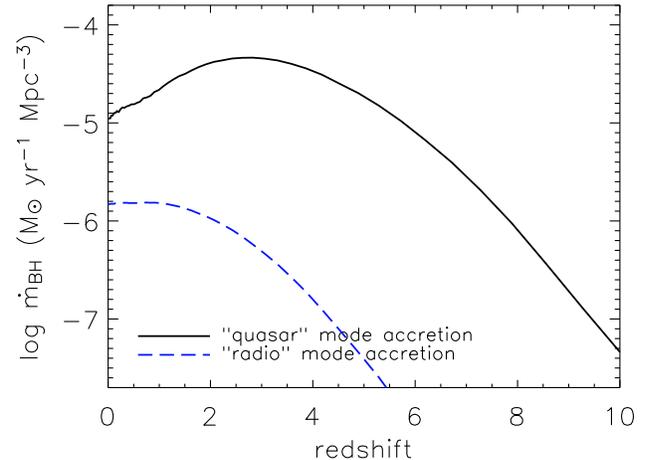}
\caption{ The black hole accretion rate density, $\dot{m}_{\rm BH}$, as a
  function of redshift for both the `quasar' and the `radio' modes discussed
  in Section~\ref{blackhole}.  This figure shows that the growth of black
  holes is dominated by the `quasar mode' at high redshift and falls off
  sharply at $z \simlt 2$.  In contrast, the `radio mode' becomes important at
  low redshifts where it suppresses cooling flows, but is not a significant
  contributor to the overall black hole mass budget.  }
\label{BHaccretion}
\end{figure}

\subsubsection{The growth of supermassive black holes}
\label{BHgrowth}
Fig.~\ref{BHaccretion} shows the evolution of the mean black hole accretion
rate per unit volume averaged over the entire Millennium Simulation.  We
separate the accretion into the `quasar' and `radio' modes described above
(solid and dashed lines respectively).  Black hole mass growth in our model is
dominated by the merger-driven `quasar mode', which is most efficient at
redshifts of two to four, dropping by a factor of five by redshift zero.  This
behaviour has similar form to but is weaker than the observed evolution with
redshift of the bright quasar population \citep[e.g.][]{Hartwick1990}. (See
also the discussion in \citealt{Kauffmann2000}).  In contrast, our `radio mode'
is significant only at late times, as expected from the scaling discussed
above, and for the high feedback efficiency assumed in Eq.~\ref{energyR}
it contributes only 5\% of the final black hole mass density. We will show,
however, that the outflows generated by this accretion can have a major impact
on the final galaxy properties.  Finally, integrating the accretion rate
density over time gives a present day black hole mass density of $3 \times
10^5 M_{\odot}$Mpc$^{-3}$, consistent with recent observational estimates
\citep{Yu2002, Merloni2004}.

\begin{figure}
\plotone{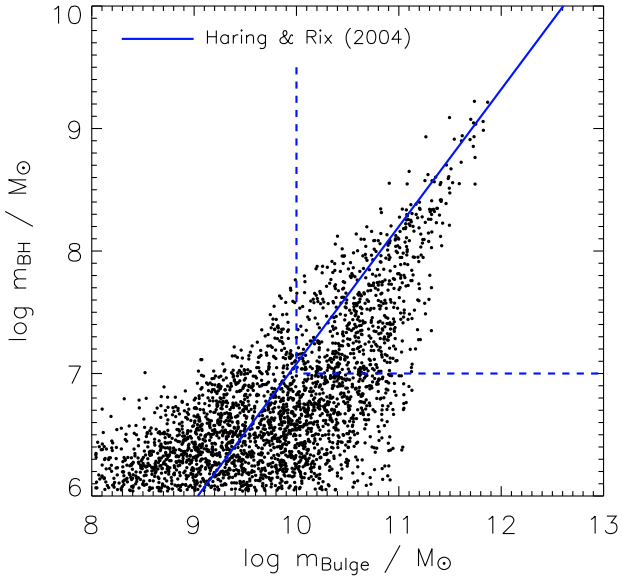}
\caption{The black hole-bulge mass relation for model galaxies at the
  present day. The local observational result of \citet{Haring2004} is
  given by the solid line, where the dashed box shows the
  approximate range over which their fit was obtained. }
\label{BHbulge}
\end{figure}

The relationship between black hole mass and bulge mass is plotted in
Fig.~\ref{BHbulge} for the local galaxy population in our `best' model.  In
this figure, the solid line shows the best fit to the observations given by
\cite{Haring2004} for a sample of $30$ nearby galaxies with well measured
bulge and black hole masses.  Their results only probe masses over the range
bounded by the dashed lines.  Our model galaxies produces a good
match to these observations with comparable scatter in the observed range
(see their Fig.~2).

\subsection{Star formation}
\label{SF}

\begin{figure}
\plotone{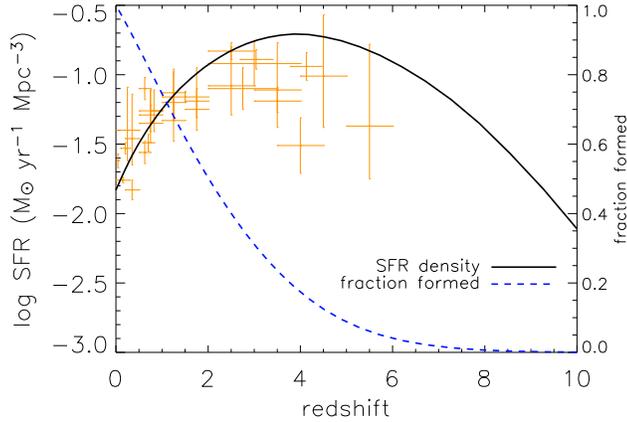}
\caption{The star formation rate density of the universe as a function of
  redshift.  The symbols show a compilation of observational results taken
  from Fig.~12 of \citet{Springel2003}.  The solid line shows our `best'
  model, which predicts that galaxies form much of their mass relatively
  early. The dashed line (and right axis) indicate the increase in stellar
  mass with redshift. Approximately $50\%$ of all stars form by $z\!=\!2$.}
\label{sfhistory}
\end{figure}

We use a simple model for star formation similar to those adopted by earlier
authors. We assume that all star formation occurs in cold disk gas, either
quiescently or in a burst (see Section~\ref{mergers}). Based on the
observational work of \cite{Kennicutt1998}, we adopt a threshold surface density
for the cold gas below which no stars form, but above which gas starts to
collapse and form stars.  According to \cite{Kauffmann1996}, this critical
surface density at a distance $R$ from the galaxy centre may be
approximated by 
\begin{equation}
\Sigma_{\rm crit} (R) = 120\, \Big( \frac{V_{\rm vir}}{200\,{\rm km\,s^{-1}}}
\Big) \Big( \frac{R}{{\rm kpc}} \Big)^{-1}\,\rm{M_{\odot} pc^{-2}} ~,  
\label{sigma_crit}
\end{equation}
We convert this critical surface density into a critical mass by assuming the
cold gas mass to be evenly distributed over the disk.  The resulting critical
cold gas mass is:
\begin{equation}
m_{\rm crit} = 3.8 \times 10^{9}\, \Big( \frac{V_{\rm vir}}{200\,{\rm km\,s^{-1}}}
\Big) \Big( \frac{r_{\rm disk}}{10\,{\rm kpc}} \Big)\, {\rm M_{\odot}} ~,  
\label{m_crit}
\end{equation}
where we assume the disk scale length to be $r_s = (\lambda /
\sqrt{2}) R_{\rm vir}$ \citep{Mo1998}, and set the outer disk radius
to $r_{\rm disk} = 3 r_s$, based on the properties of the Milky Way
\citep{vandenBergh2000}.  Here $\lambda$ is the spin parameter of the
dark halo in which the galaxy resides \citep{Bullock2001}, measured
directly from the simulation at each time step. When the mass of cold
gas in a galaxy is greater than this critical value we assume the star
formation rate to be
\begin{equation}
\dot{m}_* = \alpha_{\rm{SF}} \ (m_{\rm cold} - m_{\rm crit})\ /\
t_{\rm dyn,\rm disk}~,
\label{sfr}
\end{equation}
where the efficiency
$\alpha_{\rm{SF}}$ is typically set so that $5$ to $15$ percent of the gas
is converted into stars in a disk dynamical time $t_{\rm dyn,{\rm disk}}$, which
we define to be $r_{\rm disk}/V_{\rm vir}$.  This star formation model
produces a global star formation history consistent with the observed
star formation density of the universe out to at least $z\!=\!2$,
as shown in Fig.~\ref{sfhistory} (note that this figure also includes star
formation through starbursts -- see Section~\ref{mergers}) .

\begin{figure}
\plotone{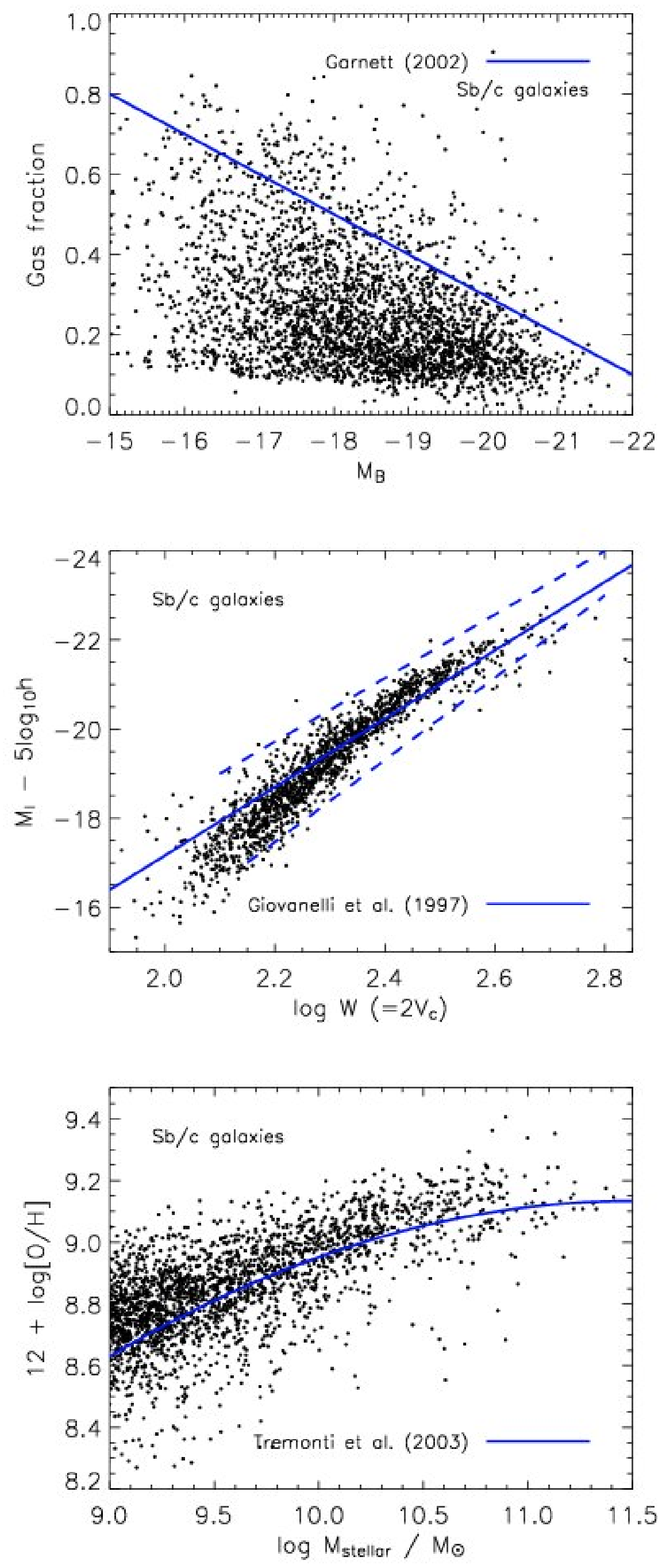}
\caption{Selected results for Sb/c galaxies (identified by bulge-to-total
  luminosity, see Section~\ref{SF}) for our best model.  (top) Gas fractions
  as a function of B-band magnitude.  The solid line is a representation of
  the mean behaviour in the (incomplete) sample of \citet{Garnett2002}.
  (middle) The Tully-Fisher relation, where the disk circular velocity, $V_c$,
  is approximated by $V_{\rm vir}$ for the dark halo.  The solid line with
  surrounding dashed lines represents the mean result and scatter found by
  \citet{Giovanelli1997}.  (bottom) Cold gas metallicity as a function of
  total stellar mass.  The solid line represents the result of
  \citet{Tremonti2004}.}
\label{spirals}
\end{figure}

When implemented in our model, Eq.~\ref{sfr} leads to episodic star formation
that self-regulates so as to maintain the critical surface density of
Eq.~\ref{sigma_crit}.  This naturally reproduces the observed spiral galaxy
gas fractions without the need for additional parameterisation, as we
demonstrate in the top panel of Fig.~\ref{spirals} using model Sb/c galaxies
identified as objects with bulge-to-total luminosity: $1.5 \le M_{\rm
I,bulge}-M_{\rm I,total} \le 2.5$ (bulge formation is described in
Section~\ref{mergers}).

\subsection{Supernova feedback}
\label{SN}

As star formation proceeds, newly formed massive stars rapidly complete their
evolution and end their life as supernovae.  These events inject gas, metals
and energy into the surrounding medium, reheating cold disk gas and possibly
ejecting gas even from the surrounding halo.

The observations of \cite{Martin1999} suggest modelling the amount of cold
gas reheated by supernovae as
\begin{equation}
\Delta m_{\rm reheated} = \epsilon_{\rm{disk}} \Delta m_*~,
\label{reheated}
\end{equation}
where $\Delta m_*$ is the mass of stars formed over some finite time interval
and $\epsilon_{\rm{disk}}$ is a parameter which we fix at
$\epsilon_{\rm{disk}}=3.5$ based on the observational data.  The energy
released in this interval can be approximated by
\begin{equation}
\Delta E_{\rm SN} = 0.5~\epsilon_{\rm{halo}}~\Delta m_* V_{\rm SN}^2~,
\label{SN_energy}
\end{equation}
where $0.5~V_{\rm SN}^2$ is the mean energy in supernova ejecta per unit mass
of stars formed, and $ \epsilon_{\rm{halo}}$ parametrises the efficiency with
which this energy is able to reheat disk gas. Based on a standard initial
stellar mass function and standard supernova theory we take $V_{\rm SN} =
630\,{\rm km\, s}^{-1}$. In addition, for our `best' model we adopt
$\epsilon_{\rm{halo}}=0.35$. If the reheated gas were added to the hot halo
without changing its specific energy, its total thermal energy would change by
\begin{equation}
\Delta E_{\rm hot} = 0.5~\Delta m_{\rm reheated} V_{\rm vir}^2~.
\label{deltaEhot}
\end{equation}
Thus the \emph{excess} energy in the hot halo after reheating is just $\Delta
E_{\rm excess} = \Delta E_{\rm SN} - \Delta E_{\rm hot}$.  When $\Delta E_{\rm
excess}\!<\!0$ the energy transferred with the reheated gas is insufficient to
eject any gas out of the halo and we assume all hot gas remains associated
with the halo.  When excess energy is present, i.e. $\Delta E_{\rm excess}\!>\!0$,
we assume that some of the hot gas is ejected from the halo into an external
`reservoir'. Specifically, we take
\begin{equation}
\Delta m_{\rm ejected} = \frac{\Delta E_{\rm excess}}{E_{\rm hot}} m_{\rm
hot} = \Big( \epsilon_{\rm halo} \frac{V_{\rm SN}^2}{V_{\rm vir}^2} -
\epsilon_{\rm disk} \Big)
\Delta m_*\, ,
\label{ejected}
\end{equation}
where $E_{\rm hot}= 0.5 \ m_{\rm hot} V_{\rm vir}^2$ is the total
thermal energy of the hot gas, and we set $\Delta m_{\rm ejected} = 0$
when this equation gives negative values (implying $\Delta E_{\rm
excess}\!<\!0$ as discussed above).  This is similar to the traditional
semi-analytic feedback recipe, $\Delta m_{\rm ejected} \propto \Delta
m_*/V_{\rm vir}^2$, but with a few additions.  For small $V_{\rm vir}$
the entire hot halo can be ejected and then $\Delta m_{\rm ejected}$
must saturate at $\Delta m_{\rm reheated}$. Conversely, no hot gas can
be ejected from the halo for $V_{\rm vir}^2 >\epsilon_{\rm halo}
V_{\rm SN}^2/\epsilon_{\rm disk}$, i.e. for halo circular velocities
exceeding about $200\,{\rm km\, s}^{-1}$ for our favoured parameters.

Ejected gas leaves the galaxy and its current halo in a wind or `super-wind',
but it need not be lost permanently.  As the dark halo grows, some of the
surrounding ejecta may fall back in and be reincorporated into the cooling
cycle.  We follow \cite{Springel2001} and \cite{deLucia2004} and model this by
assuming
\begin{equation}
\dot{m}_{\rm ejected} = -\gamma_{\rm{ej}} \,m_{\rm ejected}\,/\,t_{\rm dyn} ~,
\label{reincorporated}
\end{equation}
where $\gamma_{\rm{ej}}$ controls the amount of reincorporation per dynamical
time; typically we take $\gamma_{\rm{ej}} = 0.3$ to $1.0$.  Such values imply
that all the ejected gas will return to the hot halo in a few halo dynamical
times.

The prescriptions given in this section are simple, as well as
physically and energetically plausible, but they have little detailed
justification either from observation or from numerical simulation.
They allow us to track in a consistent way the exchange of each halo's
baryons between our four phases (stars, cold disk gas, hot halo gas,
ejecta), but should be regarded as a rough attempt to model the
complex astrophysics of feedback which will surely need significant
modification as the observational phenomenology of these processes is
explored in more depth.

Supernova feedback and star formation act together to regulate the
stellar growth of a galaxy.  This is especially important for $L<L^*$
galaxies, where feedback can eject most of the baryons from the
system, reducing the supply of star-forming material for time periods
much longer than the cooling/supernova heating cycle.  In the middle
panel of Fig.~\ref{spirals} we plot the Tully-Fisher relation for
model Sb/c galaxies (see Section~\ref{SF}).  The Tully-Fisher relation
is strongly influenced by the link between star formation and
supernova heating.  The circular velocity of a galactic disk is (to
first order) proportional to the virial velocity of the host dark
matter halo and thus to its escape velocity.  In our model (and most
others) this is closely related to the ability of the galaxy to blow a
wind. The luminosity of a galaxy is determined by its ability to turn
its associated baryons into stars. The overall efficiency of this
process in the face of supernova and AGN feedback sets the amplitude
of the Tully-Fisher relation, while the way in which the efficiency
varies between systems of different circular velocity has a strong
influence on the slope.

To predict a Tully-Fisher relation for our model we need to assign a
maximum rotation velocity to each of our galaxy disks. For central
galaxies we simply take this velocity to be the $V_{\rm vir}$ of the
surrounding halo, while for satellite galaxies we take it to be the
$V_{\rm vir}$ of the surrounding halo at the last time the galaxy was
a central galaxy. This is a crude approximation, and for realistic
halo structures it is likely to be an underestimate both because of
the concentration of the dark matter distribution and because of the
gravitational effects of the baryons \citep[e.g.][]{Mo1998}. Obtaining
a good simultaneous fit to observed luminosity functions and
Tully-Fisher relations remains a difficult problem within the
$\Lambda$CDM paradigm \citep[see, for example][]{Cole2000}. Our
unrealistic assumption for the disk rotation velocity actually
produces quite a good fit to the observational data of
\cite{Giovanelli1997}, demonstrating that our simple star formation
and feedback recipes can adequately represent the growth of stellar
mass across a wide range of scales.  We find clear deviations from
power law behaviour for $\log W \simlt 2.3$ (approximately $V_c \simlt
100\,{\rm km\,s^{-1}}$), where the efficiency of removing gas from low
mass systems combines with our threshold for the onset of star
formation to reduce the number of stars that can form. The resulting
downward bend is qualitatively similar to that pointed out in real
data by \cite{McGaugh2000}. These authors show that including the
gaseous component to construct a `baryonic' Tully-Fisher relation
brings the observed points much closer to a power-law, and the same is
true in the model we present here. Limiting star formation in galaxies
that inhabit shallow potentials has a strong effect on the faint-end
of the galaxy luminosity function, as will be seen in
Section.~\ref{LumDist}.

\subsection{Galaxy morphology, merging and starbursts}
\label{mergers}

In the model we discuss here, the morphology of a galaxy is assumed to
depend only on its bulge-to-total luminosity ratio, which in turn is
determined by three distinct physical processes: disk growth by
accretion, disk buckling to produce bulges, and bulge formation
through mergers.  We treat disk instabilities using the simple
analytic stability criterion of \cite{Mo1998}; the stellar disk of a
galaxy becomes unstable when the following inequality is met,
\begin{equation}
\frac{V_c}{({\rm G} m_{\rm{D}} / r_{\rm{D}})^{1/2}} \le 1 ~,
\label{disk_stability}
\end{equation}
where we again approximate the rotation velocity of the disk $V_{\rm
c}$ by $V_{\rm vir}$.  For each galaxy at each time-step we evaluate
the left-hand side of Eq.~\ref{disk_stability}, and if it is smaller
than unity we transfer enough stellar mass from disk to bulge (at
fixed $r_{\rm{D}}$) to restore stability.

Galaxy mergers shape the evolution of galaxies, affecting both their
morphology and (through induced starbursts) their star formation
history. Mergers can occur in our model between the central galaxy of
a dark halo or subhalo and a satellite galaxy which has lost its own
dark subhalo.  Substructure is followed in the Millennium Run down to
a 20 particle limit, which means that the orbit of a satellite galaxy
within a larger halo is followed explicitly until its subhalo mass
drops below $1.7\times 10^{10}h^{-1}M_\odot$. After this point, the
satellite's position and velocity are represented by those of the most
bound particle of the subhalo at the last time it was identified. At
the same time, however, we start a merger `clock' and estimate a
merging time for the galaxy using the dynamical friction formula of
\cite{Binney1987},
\begin{equation}
t_{\rm friction} = 1.17 \frac{V_{\rm vir} r_{\rm sat}^2}{{\rm G}
  m_{\rm sat} \ln \Lambda} ~. 
\label{merging_time}
\end{equation}
This formula is valid for a satellite of mass $m_{\rm sat}$ orbiting
in an isothermal potential of circular velocity $V_{\rm vir}$ at
radius $r_{\rm sat}$. We take $m_{\rm sat}$ and $r_{\rm sat}$ to be
the values measured for the galaxy at the last time its subhalo could
be identified.  The Coulomb logarithm is approximated by $\ln \Lambda
= \ln (1+M_{\rm vir}/m_{\rm sat})$.  The satellite is then merged with
the central galaxy a time $t_{\rm friction}$ after its own subhalo was
last identified. If the main halo merges with a larger system before
this occurs, a new value for $t_{\rm friction}$ is calculated and the
merger clock is restarted.

\begin{figure*}
\plotfull{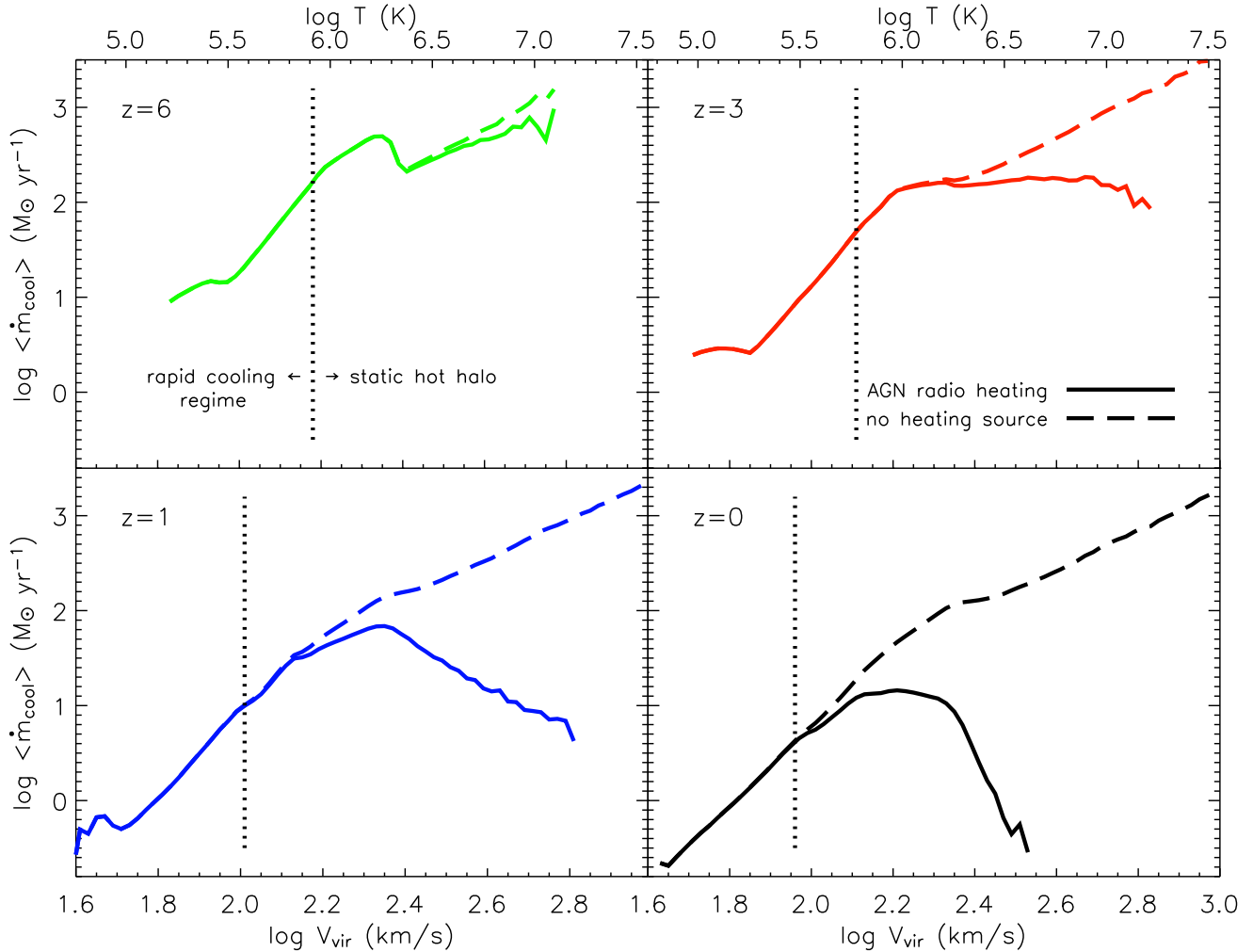}
\caption{ The mean condensation rate, $\langle\dot{m}_{\rm cool}\rangle$ as a
  function of halo virial velocity $V_{\rm vir}$ at redshifts of $6$, $3$,
  $1$, and $0$.  Solid and dashed lines in each panel represent the
  condensation rate with and without `radio mode' feedback respectively, while
  the vertical dotted lines show the transition between the rapid cooling and
  static hot halo regimes, as discussed in Section~\ref{cooling}.  This figure
  demonstrates that cooling flow suppression is most efficient in our model
  for haloes with $V_{\rm vir} > 150\,{\rm km\,s^{-1}}$ and at $z \leq 3$.}
\label{CoolingGas}
\end{figure*}

The outcome of the merger will depend on the baryonic mass ratio of
the two progenitors .  When one dominates the process, i.e. a small
satellite merging with a larger central galaxy, the stars of the
satellite are added to the bulge of the central galaxy and a minor
merger starburst (see below) will result.  The cold gas of the 
satellite is added to the disk of the central galaxy along with any
stars that formed during the burst.  Such an event is called a
\emph{minor merger}.

If, on the other hand, the masses of the progenitors are comparable a
\emph{major merger} will result.  Under these circumstances the
starburst is more significant, with the merger destroying the disks of
both galaxies to form a spheroid in which all stars are placed.  The
dividing line between a major and minor merger is given by the
parameter $T_{\rm{merger}}$: when the mass ratio of the merging
progenitors is larger than $T_{\rm{merger}}$ a major merger results,
otherwise the event is a minor merger.  Following \cite{Springel2001}
we choose $T_{\rm{merger}}=0.3$ and keep this fixed throughout.

Our starburst implementation is based on the `collisional starburst'
model of \cite{Somerville2001}.  In this model, a fraction $e_{\rm
burst}$ of the combined cold gas from the two galaxies is turned into
stars as a result of the merger:
\begin{equation}
e_{\rm burst} = \beta_{\rm burst} (m_{\rm sat} / m_{\rm
  central})^{\alpha_{\rm burst}} ~, 
\label{e_burst}
\end{equation}
where the two parameters are taken as $\alpha_{\rm burst}=0.7$ and $\beta_{\rm
burst}=0.56$. This model provides a good fit to the numerical results of
\cite{Cox2004} and also \cite{Mihos1994, Mihos1996} for merger mass ratios
ranging from 1:10 to 1:1.

\subsection{Spectroscopic evolution and dust}
\label{spectro}

The photometric properties of our galaxies are calculated using stellar
population synthesis models from \cite{Bruzual2003}.  Our implementation is
fully described in \cite{deLucia2004} and we refer the reader there (and to
references therein) for further details.

To include the effects of dust when calculating galaxy luminosities we apply
the simple `plane-parallel slab' model of \cite{Kauffmann1999}.  This model is
clearly oversimplified, but it permits us to make a reasonable first-order
correction for dust extinction in actively star-forming galaxies.  For the
details of this model we refer the reader to \cite{Kauffmann1999} and to
references therein.

\begin{figure*}
\plotfull{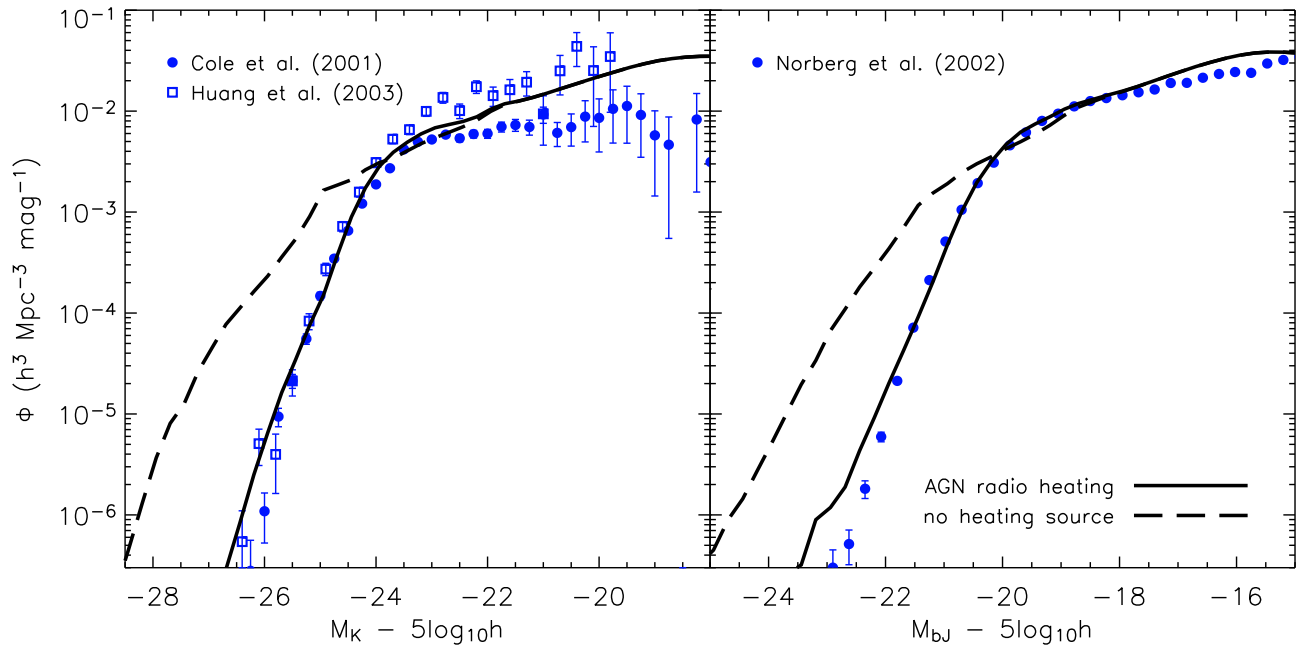}
\caption{Galaxy luminosity functions in the K (left) and ${\rm b_J}$ (right)
  photometric bands, plotted with and without `radio mode' feedback (solid and
  long dashed lines respectively -- see Section~\ref{blackhole}).  Symbols
  indicate observational results as listed in each panel.  As can be seen, the
  inclusion of AGN heating produces a good fit to the data in both colours.
  Without this heating source our model overpredicts the luminosities of
  massive galaxies by about two magnitudes and fails to reproduce the sharp
  bright end cut-offs in the observed luminosity functions.}
\label{mainLFs}
\end{figure*}

\subsection{Metal enrichment}
\label{metals}

Our treatment of metal enrichment is essentially identical to that described
in \cite{deLucia2004}.  In this model a yield $Y$ of heavy elements is
returned for each solar mass of stars formed.  These metals are produced
primarily in the supernovae which terminate the evolution of short-lived,
massive stars. In our model we deposit them directly into the cold gas in the
disk of the galaxy. (An alternative would clearly be to add some fraction of
the metals directly to the hot halo. Limited experiments suggest that this
makes little difference to our main results.)  We also assume that a fraction
$R$ of the mass of newly formed stars is recycled immediately into the cold
gas in the disk, the so called `instantaneous recycling approximation'
\citep[see][]{Cole2000}.  For full details on metal enrichment and exchange
processes in our model see \cite{deLucia2004}.  In the bottom panel of
Fig.~\ref{spirals} we show the metallicity of cold disk gas for model Sb/c
galaxies (selected, as before, by bulge-to-total luminosity, as described
in Section~\ref{SF}) as a function of total stellar mass.  For comparison, we
show the result of \cite{Tremonti2004} for mean HII region abundances in SDSS
galaxies.

\section{Results}
\label{results}

\begin{figure}
\plotone{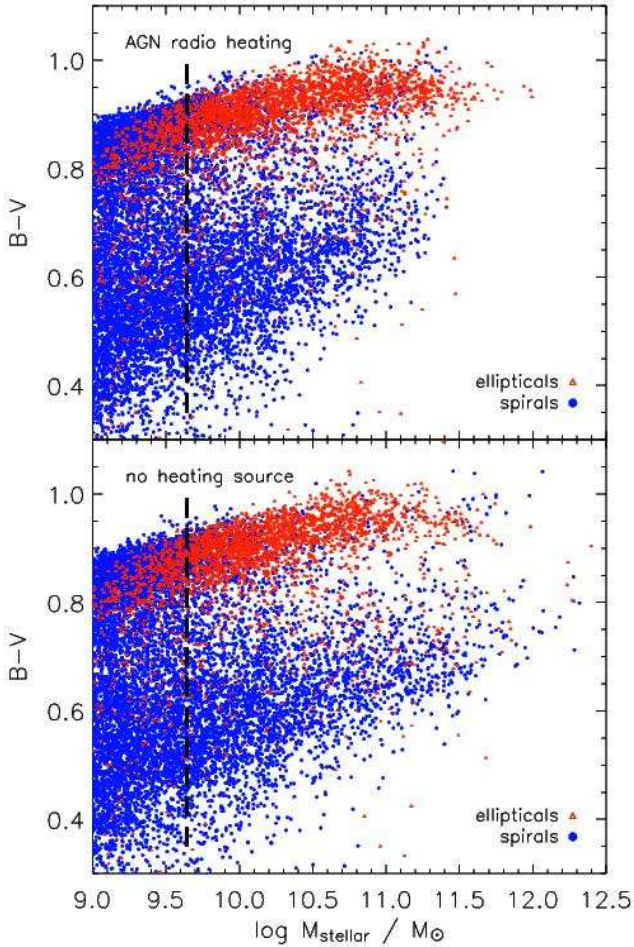}
\caption{The ${\rm B\!-\!V}$ colours of model galaxies plotted as a function
  of their stellar mass with (top) and without (bottom) `radio mode' feedback
  (see Section~\ref{blackhole}).  A clear bimodality in colour is seen in
  both panels, but without a heating source the most massive galaxies are blue
  rather than red.  Only when heating is included are massive galaxies as red
  as observed.  Triangles (red) and circles (blue) correspond to early and
  late morphological types respectively, as determined by bulge-to-total
  luminosity ratio (see Section~\ref{LumDist}).  The thick dashed lines mark
  the resolution limit to which morphology can be reliably determined in the
  Millennium Run.}
\label{BminusV}
\end{figure}

\begin{figure}
\plotone{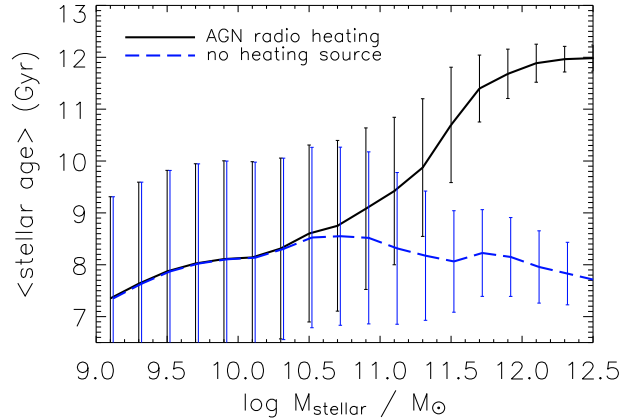}
\caption{Mean stellar ages of galaxies as a function of stellar mass for
  models with and without `radio mode' feedback (solid and dashed lines
  respectively).  Error bars show the {\it rms} scatter around the mean for
  each mass bin.  The suppression of cooling flows raises the mean age of
  high-mass galaxies to large values, corresponding to high formation
  redshifts.}
\label{meanAge}
\end{figure}

In this section we examine how the suppression of cooling flows in massive
systems affects galaxy properties.  As we will show, the effects are only
important for high mass galaxies. Throughout our analysis we use the galaxy
formation model outlined in the previous sections with the parameter choices
of Table~\ref{parameters} except where explicitly noted.

\subsection{The suppression of cooling flows}
\label{coolingflows}

We begin with Fig.~\ref{CoolingGas}, which shows how our `radio mode' heating
model modifies gas condensation.  We compare mean condensation rates with and
without the central AGN heating source as a function of halo virial velocity
(solid and dashed lines respectively).  Recall that virial velocity provides a
measure of the equilibrium temperature of the system through $T_{\rm vir}
\propto V_{\rm vir}^2$, as indicated by the scale on the top axis.  The four
panels show the behaviour at four redshifts between six and the present day.
The vertical dotted line in each panel marks haloes for which $r_{\rm cool} =
R_{\rm vir}$, the transition between systems that form static hot haloes and
those where infalling gas cools rapidly onto the central galaxy disk (see
section \ref{cooling} and Fig.~\ref{rcool}).  This transition moves to haloes
of lower temperature with time, suggesting a `down-sizing' of the
characteristic mass of actively star-forming galaxies. At lower $V_{\rm vir}$
gas continues to cool rapidly, while at higher $V_{\rm vir}$ new fuel for star
formation must come from cooling flows which are affected by `radio mode'
heating.

The effect of `radio mode' feedback is clearly substantial. Suppression of
condensation becomes increasingly effective with increasing virial temperature
and decreasing redshift. The effects are large for haloes with $V_{\rm vir}
\simgt 150\,{\rm km\,s^{-1}}$ ($T_{\rm vir}\simgt 10^6$K) at $z \simlt 3$.
Condensation stops almost completely between $z\!=\!1$ and the present in haloes
with $V_{\rm vir} > 300\,{\rm km\,s^{-1}}$ ($T_{\rm vir} > 3 \times 10^6$K).
Such systems correspond to the haloes of groups and clusters which are
typically observed to host massive elliptical or cD galaxies at their centres.
Our scheme thus produces results which are qualitatively similar to the {\it
ad hoc} suppression of cooling flows assumed in previous models of galaxy
formation.  For example, \cite{Kauffmann1999} switched off gas condensation in
all haloes with $V_{\rm vir} > 350\,{\rm km\,s^{-1}}$, while \cite{Hatton2003}
stopped condensation when the bulge mass exceeded a critical threshold.

\subsection{Galaxy properties with and without AGN heating}
\label{LumDist}

The suppression of cooling flows in our model has a dramatic effect on the
bright end of the galaxy luminosity function.  In Fig.~\ref{mainLFs} we
present ${\rm K}$- and ${\rm b_J}$-band luminosity functions (left and right
panels respectively) with and without `radio mode' feedback (solid and dashed
lines respectively).  The luminosities of bright galaxies are reduced by up to
two magnitudes when the feedback is switched on, and this induces a relatively
sharp break in the luminosity function which matches the observations well. We
demonstrate this by overplotting K-band data from \cite{Cole2001} and
\cite{Huang2003} in the left panel, and ${\rm b_J}$-band data from
\cite{Norberg2002b} in the right panel.  In both band-passes the model is
quite close to the data over the full observed range.  We comment on some of
the remaining discrepancies below.

\begin{figure*}
\plotfull{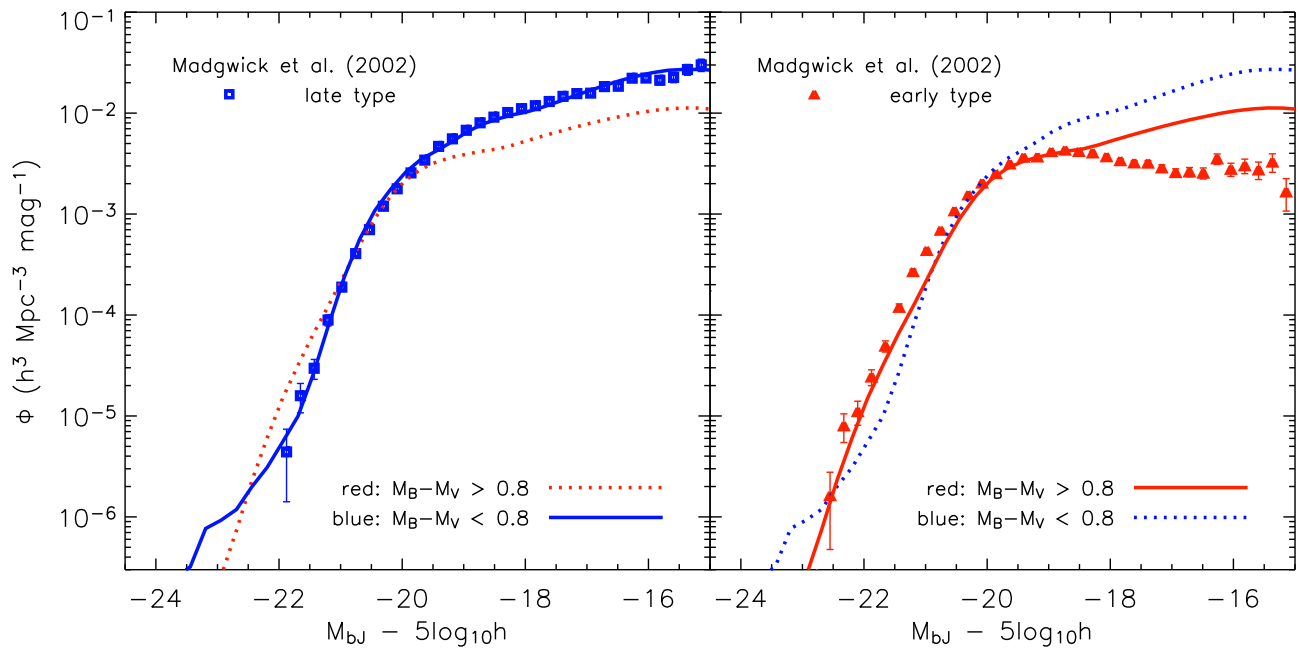}
\caption{The ${\rm b_J}$-band galaxy luminosity function split by colour at
  ${\rm B\!-\!V} =0.8$ (Fig.~\ref{BminusV}) into blue (left panel) and red
  (right panel) sub-populations (solid lines).  The dotted lines in each panel
  repeat the opposite colour luminosity function for reference.  Symbols
  indicate the observational results of \citet{Madgwick2002} for early and
  late-type 2dFGRS galaxies, split according to spectral type. Although our
  model split by colour captures the broad behaviour of the observed
  type-dependent luminosity functions, there are important differences which
  we discuss in Section~\ref{LumDist}. }
\label{typeLFs}
\end{figure*}

Our feedback model also has a significant effect on bright galaxy colours, as
we show in Fig.~\ref{BminusV}.  Here we plot the ${\rm B\!-\!V}$ colour
distribution as a function of stellar mass, with and without the central
heating source (top and bottom panels respectively).  In both panels we have
colour-coded the galaxy population by morphology as estimated from
bulge-to-total luminosity ratio (split at $L_{\rm bulge}/L_{\rm total} =
0.4$).  Our morphological resolution limit is marked by the dashed line at a
stellar mass of $\sim4\times 10^{9} M_{\odot}$; this corresponds approximately
to a halo of 100 particles in the Millennium Run.  Recall that a galaxy's
morphology depends both on its past merging history and on the stability of
its stellar disk in our model.  Both mergers and disk instabilities contribute
stars to the spheroid, as described in Section~\ref{mergers}. The build-up of
haloes containing fewer than 100  particles is not followed in enough detail
to model these processes robustly.

A number of important features can be seen in Fig.~\ref{BminusV}.  Of note is
the bimodal distribution in galaxy colours, with a well-defined red sequence
of appropriate slope separated cleanly from a broader `blue cloud'.  It is
significant that the red sequence is composed predominantly of early-type
galaxies, while the blue cloud is comprised mostly of disk-dominated systems.
This aspect of our model suggests that that the physical processes which
determine morphology (i.e. merging, disk instability) are closely related to
those which control star formation history (i.e. gas supply) and thus
determine galaxy colour.  The red and blue sequences both display a strong
metallicity gradient from low to high mass (c.f. Fig.~\ref{spirals}), and it
is this which induces a `slope' in the colour-magnitude relations which agrees
well with observation \citep[e.g.][]{Baldry2004}.

By comparing the upper and lower panels in Fig.~\ref{BminusV} we can
see how `radio mode' feedback modifies the luminosities, colours and
morphologies of high mass galaxies.  Not surprisingly, the brightest
and most massive galaxies are also the reddest and are ellipticals
when cooling flows are suppressed, whereas they are brighter, more
massive, much bluer and typically have disks if cooling flows continue
to supply new material for star formation.  AGN heating cuts off the
gas supply to the disk from the surrounding hot halo, truncating star
formation and allowing the existing stellar population to redden.
However, these massive red galaxies do continue to grow through
merging.  This mechanism allows the dominant cluster galaxies to gain
a factor 2 or 3 in mass without significant star formation, in
apparent agreement with observation \citep{Aragon1998}. This
late-stage (i.e. $z\simlt 1$) hierarchical growth moves objects to
higher mass without changing their colours.

It is also interesting to examine the effect of AGN heating on the
stellar ages of galaxies. In Fig.~\ref{meanAge} the solid and dashed
lines show mean stellar age as a function of stellar mass for models
with and without `radio mode' feedback, while error bars indicate the
{\it rms} scatter around the mean.  Substantial differences are seen
for galaxies with $M_{\rm stellar} \simgt 10^{11} M_{\odot}$: the mean
age of the most massive galaxies approaching $12$ Gyr when cooling
flows are suppressed but remaining around $8$ Gyr when feedback is
switched off.  Such young ages are clearly inconsistent with the old
stellar populations observed in the majority of massive cluster
ellipticals.

The colour bimodality in Fig.~\ref{BminusV} is so pronounced that it
is natural to divide our model galaxies into red and blue populations
and to study their properties separately. We do this by splitting at
${\rm B\!-\!V} = 0.8$, an arbitrary but natural
choice. Fig.~\ref{typeLFs} shows separate ${\rm b_J}$ luminosity
functions for the resulting populations.  For comparison we overplot
observational results from \cite{Madgwick2002} for 2dFGRS galaxies
split by spectral type.  Their luminosity functions are essentially
identical to those of \cite{Cole2005}, who split the 2dFGRS catalogue
by ${\rm b_J}$-${\rm r_F}$ colour. It thus can serve to indicate the
observational expectations for populations of different colour.  The
broad behaviour of the red and blue populations is similar in the
model and in the 2dFGRS.  The faint-end of the luminosity function is
dominated by late-types, whereas the bright end has an excess of
early-types. The two populations have equal abundance about half to
one magnitude brighter than $M_{{\rm b_J}}^*$ \citep{Norberg2002b}.

Fig.~\ref{typeLFs} also shows some substantial differences between
model and observations. The red and blue populations differ more in
the real data than they do in the model. There is a tail of very
bright blue galaxies in the model, which turn out to be objects
undergoing strong, merger-induced starbursts.  These correspond in
abundance, star formation rate and evolutionary state to the observed
population of Ultraluminous Infrared Galaxies (ULIRG's) with the
important difference that almost all the luminosity from young stars
in the real systems is absorbed by dust and re-emitted in the mid- to
far-infrared \citep{Sanders1996}. Clearly we need better dust
modelling than our simple `slab' model (Section~\ref{spectro}) in
order to reproduce the properties of such systems adequately.  If we
suppress starbursts in bright galaxy mergers we find that the blue
tail disappears and the observed behaviour is recovered.  A second and
substantial discrepancy is the apparent overproduction of faint red
galaxies in our model, as compared to the 2dF measurements
\citep[however see][]{Popesso2005, Gonzalez2005}.  Further work is
clearly needed to understand the extent and significance of this
difference.

\section{Physical models of AGN feedback}
\label{discussion}

Our phenomenological model for `radio mode' feedback (Section~\ref{radio}) is
not grounded in any specific model for hot gas accretion onto a black hole or
for the subsequent generation and thermalization of energy through radio
outflows. Rather it is based on the observed properties of cooling flows and
their central radio sources, and on the need for a source of feedback which
can suppress gas condensation onto massive galaxies without requiring the
formation of new stars. We have so far focused on the effects of such
feedback without discussing how it might be realised.  In this section we
present two physical models which suggest how accretion onto the central
black hole may lead to activity in a way which could justify the parameter
scalings we have adopted.

\subsection{Cold cloud accretion}
\label{coldclouds}

A simple picture for cooling flow evolution, based on the similarity
solution of \cite{Bertschinger1989} for an unperturbed halo in
isolation, can be summarised as follows.  Cooling flows develop in any
halo where the cooling time of the bulk of the hot gas is longer than
the age of the system so that a static hot halo can form. Such haloes
usually have a strong central concentration and we approximate their
structure by a singular isothermal sphere. The inner regions then have
a local cooling time shorter than the age of the system, and the gas
they contain radiates its binding energy and flows inwards. The flow
region is bounded by the cooling radius $r_{\rm cool}$ where the local
cooling time is equal to the age of the system (see
section~\ref{cooling}).  This radius increases with time as
$t^{1/2}$. As Bertschinger showed, the temperature of the gas {\it
increases} by about 20\% as it starts to flow inwards, and its density
profile flattens to $\rho_g\propto r^{-3/2}$. Initially, the flow is
subsonic and each gas shell sinks stably and isothermally in
approximate hydrostatic equilibrium.  As it sinks, however, its inward
velocity accelerates because its cooling time shrinks more rapidly
than the sound travel time across it, and at the sonic radius, $r_{\rm
sonic}$, the two become equal. At this point the shell goes into free
fall, its temperature decreases rapidly and it may fragment as a
result of thermal instability \citep{Cowie1980, Nulsen1986,
Balbus1989}. The dominant component of the infalling gas is then in
the form of cold clouds and is no longer self-coupled by hydrodynamic
forces. Different clouds pursue independent orbits, some with
pericentres perhaps orders of magnitude smaller than $r_{\rm
sonic}$. If these lie within the zone of influence of the black hole,
$r_{\rm BH} = {\rm G} m_{\rm BH}/V_{\rm vir}^2$, we assume that some
of the cold gas becomes available for fuelling the radio source;
otherwise we assume it to be added to the cold gas disk.
 
The parameter scalings implied by this picture can be estimated as
follows.  The sound travel time across a shell at the cooling radius
is shorter than the cooling time by a factor $\sim \, r_{\rm
cool}/R_{\rm vir}$. At smaller radii the ratio of cooling time to
sound travel time decreases as $r^{1/2}$ so that $r_{\rm sonic}/r_{\rm
cool} \sim (r_{\rm cool}/R_{\rm vir})^2$ implying $r_{\rm sonic} \sim
\, r_{\rm cool}^3/R_{\rm vir}^2$.  If we adopt $r_{\rm BH} > 10^{-4}
\, r_{\rm sonic}$ as the condition for effective fuelling of the radio
source, we obtain
\begin{equation}
m_{\rm BH} > 10^{-4}\, M_{\rm vir}\, (r_{\rm cool}/R_{\rm vir})^3
\label{sonic}
\end{equation}
as the corresponding minimum black hole mass for fragmented clouds to
be captured.  Under such conditions, only a small fraction ($\sim
0.01\%$) of the cooling flow mass need be accreted to halt the flow.
The ratio in parentheses on the right-hand side of this equation
scales approximately as $r_{\rm cool}/R_{\rm vir}\propto (m_{\rm
hot}/M_{\rm vir})^{1/2}t_{\rm H}^{-1/2} V_{\rm vir}^{-1}$, so the minimum
black hole mass scales approximately as $(m_{\rm hot}/M_{\rm
vir})^{3/2}t_{\rm H}^{-1/2}$ and is almost independent of $V_{\rm vir}$. In
our model, the growth of black holes through mergers and `quasar mode'
accretion produces a population where mass increases with time and
with host halo mass. As a result, effective fuelling takes place
primarily in the more massive haloes and at late times for this `cold
cloud' prescription.
  
\begin{figure}
\plotone{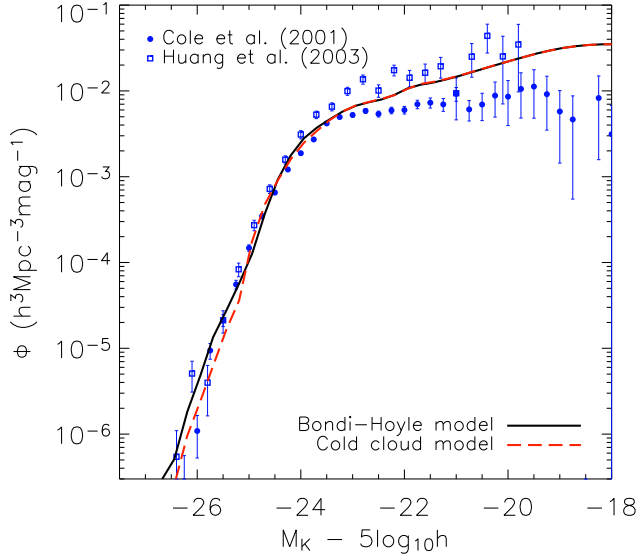}
\caption{The observed K-band galaxy luminosity function is compared with the
  results from models using the two physical prescriptions for `radio mode'
  accretion discussed in Section~\ref{discussion}: the Bondi-Hoyle accretion
  model (solid line) and the cold cloud accretion model (dashed line).
  Symbols indicate observational data from \citet{Cole2001} and
  \citet{Huang2003}.  Both models can produce a luminosity function which 
  matches observation well.}
\label{physical}
\end{figure}

To test this particular model we switch off our standard phenomenological
treatment of `radio mode' feedback (section~\ref{radio}), assuming instead
that feedback occurs only when Eq.~\ref{sonic} is satisfied and that in
this case it is sufficient to prevent further condensation of gas from the
cooling flow.  All other elements of our galaxy and black hole formation model
are unchanged. The resulting cooling flow suppression is similar to that seen
in Fig.~\ref{CoolingGas}, and all results presented in Section~\ref{sam} and
\ref{results} are recovered.  An illustration of this is given by
Fig.~\ref{physical}, where we compare the K-band luminosity function from this
particular model (the dashed line) to the observational data (c.f. also
Fig.~\ref{mainLFs}).  The model works so well, of course, because the
numerical coefficient in  Eq.~\ref{sonic} is uncertain and we have taken
advantage of this to choose a value which puts the break in the luminosity
function at the observed position. This adjustment plays the role of the
efficiency parameter $\kappa_{\rm AGN}$ in our standard analysis (see
Eq.~\ref{accretionR}).

\subsection{Bondi-Hoyle accretion}
\label{bondi}

Our second physical model differs from the first in assuming that accretion is
not from the dominant, cold cloud component which forms within the sonic
radius, but rather from a subdominant hot component which fills the space
between these clouds. The clouds themselves are assumed to be lost to the
star-forming disk. The density profile of the residual hot component was
estimated by \citet{Nulsen2000} from the condition that the cooling time of
each radial shell should remain equal to the sound travel time across it as it
flows inwards. This requires the density of the hot component to vary
as $1/r$ within $r_{\rm sonic}$ and thermal instabilities must continually
convert material into condensed clouds in order to maintain this structure
as the hot gas flows in.

The rate at which hot gas is accreted onto the black hole can then be
estimated from the Bondi-Hoyle formula \citep{Bondi1952, Edgar2004}:
\begin{equation}
\dot{m}_{\rm Bondi} = 2.5 \pi {\rm G}^2 \frac{m_{\rm BH}^2
  \rho_0}{c_{\rm s}^3}~. 
\label{physicalBH}
\end{equation}
Here $\rho_0$ is the (assumed uniform) density of hot gas around the black
hole, and in all that follows we approximate the sound speed, $c_{\rm s}$, by
the virial velocity of the halo, $V_{\rm vir}$. Of course, the density
distribution of gas surrounding the black hole is {\it not} uniform so the
question immediately arises as to what density we should choose.  We follow a
suggestion of E. Churazov and use the value predicted by the `maximum cooling
flow' model of \citet{Nulsen2000} at the Bondi radius $r_{Bondi} \equiv 2GM_{\rm
BH}/c_{\rm s}^2 = 2 r_{\rm BH}$, the conventional boundary of the sphere of
influence of the black hole. We therefore equate the sound travel time across
a shell at this radius to the local cooling time there:
 \begin{equation}
\frac{2r_{\rm Bondi}}{c_{\rm s}} \approx \frac{4 {\rm G} m_{\rm BH}}{V_{\rm
 vir}^3} = \frac{3}{2} \frac{\bar{\mu} m_p k T }{ \rho_g (r_{\rm Bondi})
 \Lambda(T,Z)}~.
\label{timescaleBH}
\end{equation}
Solving for the density gives
\begin{equation}
\rho_0 = \rho_g(r_{\rm Bondi}) = \frac{3 \mu m_p}{8 {\rm G}}
\frac{kT}{\Lambda} \frac{V_{\rm vir}^3}{m_{\rm BH}}~.
\label{densityBH}
\end{equation}
Combining Eq.~\ref{densityBH} with \ref{physicalBH} provides us with
the desired estimate for the hot gas accretion rate onto the black hole:
\begin{equation}
\dot{m}_{\rm Bondi} \approx {\rm G} \mu m_p \frac{kT}{\Lambda} m_{\rm BH}~.
\label{accretionBH}
\end{equation}
Notice that this rate depends only on the black hole mass and on the virial
temperature of the halo. It is independent both of time and of $m_{\rm
hot}/M_{\rm vir}$, the hot gas fraction of the halo. It is valid as long as
$r_{\rm Bondi} < r_{\rm sonic}$, which is always the case in our models.

To investigate the effects of this model we replace the
phenomenological `radio mode' accretion rate of Eq.~\ref{accretionR}
with that given by Eq.~\ref{accretionBH}.  Since the latter has no
adjustable efficiency, we use the energy generation parameter $\eta$
of Eq.~\ref{energyR} to control the effectiveness of cooling flow
suppression.  (This was not necessary before since $\eta$ always
appeared in the product $\eta\, \kappa_{\rm AGN}$, where $\kappa_{\rm
AGN}$ is the efficiency parameter of Eq.~\ref{accretionR}.)  With this
change of cooling flow accretion model and taking $\eta = 0.03$, we
are able to recover the results of Sections~\ref{sam} and
\ref{results} without changing any other aspects of our galaxy and
black hole formation model.  The final galaxy population is, in fact,
almost identical to that presented in previous sections. This is not
surprising, perhaps, since Eq.~\ref{accretionBH} has very similar
scaling properties to Eq.~\ref{accretionR}.  In Fig.~\ref{physical} we
illustrate the success of the model by overplotting its prediction for
the K-band luminosity function (the solid line) on the observational
data and on the prediction of the cold cloud accretion model of the
last subsection. The two models agree very closely both with each
other and with our standard phenomenological model (see
Fig.~\ref{mainLFs}).

\section{Conclusions}
\label{conclusion}
AGN feedback is an important but relatively little explored element in the
co-evolution of galaxies and the supermassive black holes at their centres. In
this paper we set up machinery to study this co-evolution in unprecedented
detail using the very large Millennium Run, a high-resolution simulation of
the growth of structure in a representative region of the concordance
$\Lambda$CDM cosmology. Most of our modelling follows earlier work, but in an
important extension we introduce a `radio' feedback mode, based on simple
physical models and on the observed phenomenology of radio sources in cooling
flows.  This mode suppresses gas condensation at the centres of massive haloes
without requiring the formation of new stars.  Our modelling produces large
catalogues of galaxies and supermassive black holes which can be used to
address a very wide range of issues concerning the evolution and clustering of
galaxies and AGN.  Some clustering results were already presented in
\citet{Springel2005}. In the present paper, however, we limit ourselves to
presenting the model in some detail and to investigating the quite dramatic
effects which `radio mode' feedback can have on the luminosities and colours
of massive galaxies. Our main results can be summarised as follows:

\begin{itemize}

\item[(i)] We study the amount of gas supplied to galaxies in each of
  the two gas infall modes discussed by \cite{White1991}: the `static
  halo' mode where postshock cooling is slow and a quasistatic hot
  atmosphere forms behind the accretion shock; and the `rapid cooling'
  mode where the accretion shock is radiative and no such atmosphere
  is present.  We distinguish these two modes using the criterion of
  \cite{White1991} as modified by \cite{Springel2001} and tested
  explicitly using SPH simulations by \cite{Yoshida2002}.  Our results
  show a sharp transition between the two regimes at a halo mass of
  $2$--$3 \times 10^{11} M_{\odot}$. This division depends on the
  chemical enrichment prescription adopted and moves from higher to
  lower $V_{\rm vir}$ with time (corresponding to approximately
  constant $M_{\rm vir}$), suggesting a `down-sizing' of star
  formation activity as the bulk of the gas accreted by the haloes of
  larger systems is no longer available in the interstellar medium of
  the central galaxy.

\indent \item[(ii)] We have built a detailed model for cooling, star
  formation, supernova feedback, galaxy mergers and metal enrichment based on
  the earlier models of \cite{Kauffmann1999}, \cite{Springel2001} and
  \cite{deLucia2004}. Applied to the Millennium Run this model reproduces many
  of the observed properties of the local galaxy population: the Tully-Fisher,
  cold gas fraction/stellar mass and cold gas metallicity/stellar mass
  relations for Sb/c spirals (Fig.~\ref{spirals}); the field galaxy luminosity
  functions (Fig.~\ref{mainLFs} \& \ref{typeLFs}); the colour-magnitude
  distribution of galaxies (Fig.~\ref{BminusV}); and the increase in mean
  stellar age with galaxy mass (Fig.~\ref{meanAge}). In addition the model
  produces a global star formation history in reasonable agreement with
  observation (Fig.~\ref{sfhistory}). We also show in \citet{Springel2005}
  that the $z\!=\!0$ clustering properties of this population are in good
  agreement with observations.

\indent \item[(iii)] Our black hole implementation extends the previous work
  of \cite{Kauffmann2000} by assuming three modes of AGN growth: merger-driven
  accretion of cold disk gas in a `quasar mode', merging between black holes,
  and `radio mode' accretion when a massive black hole finds itself at the
  centre of a static hot gas halo.  The `quasar mode' is the dominant source
  for new black hole mass and is most active between redshifts of four and
  two. The `radio mode' grows in overall importance until $z\!=\!0$ and is
  responsible for the feedback which shuts off the gas supply in cooling
  flows.  This model reproduces the black hole mass/bulge mass relation
  observed in local galaxies (Fig.~\ref{BHbulge}). The global history of
  accretion in the `quasar mode' is qualitatively consistent with the
  evolution inferred from the optical AGN population (Fig.~\ref{BHaccretion}).

\indent \item[(iv)] Although the overall accretion rate is low, `radio mode'
  outflows can efficiently suppress condensation in massive systems
  (Fig.~\ref{CoolingGas}). As noted by many authors who have studied the
  problem in more detail than we do, this provides an energetically feasible
  solution to the long-standing cooling flow `problem'.  Our analysis shows
  that the resulting suppression of gas condensation and star formation can
  produce luminosity functions with very similar bright end cut-offs to those
  observed (Fig.~\ref{mainLFs}), as well as colour-magnitude distributions in
  which the most massive galaxies are red, old and elliptical, rather
  than blue, young and disk-dominated (Figs~\ref{BminusV}
  and~\ref{meanAge}). 

\indent \item[(v)] The ${\rm B\!-\!V}$ colour distribution of galaxies is
  bimodal at all galaxy masses. Galaxies with early-type bulge-to-disk ratios
  are confined to the red sequence, as are the most massive galaxies, and the
  most massive galaxies are almost all bulge-dominated, as observed in the
  real universe (Fig.~\ref{BminusV}).  This bimodality provides a natural
  division of model galaxies into red and blue subpopulations.  The
  colour-dependent luminosity functions are qualitatively similar to those
  found for early and late-type galaxies in the 2dFGRS (Fig.~\ref{typeLFs}),
  although there are significant discrepancies.  After exhausting their cold
  gas, massive central galaxies grow on the red sequence through `burstless'
  merging, gaining a factor of two or three in mass without significant star
  formation \citep{Aragon1998}.  Such hierarchical growth does not
  change a galaxy's colour significantly, moving it brightward almost
  parallel to the colour-magnitude relation.

\indent \item[(vi)] We present two physical models for black hole accretion
  from cooling flow atmospheres. We suppose that this accretion is
  responsible for powering the radio outflows seen at the centre of
  almost all real cooling flows.  The models differ in their
  assumptions about how gas accretes from the inner regions of the
  cooling flow, where it is thermally unstable and dynamically
  collapsing. One assumes accretion of cold gas clouds if these come
  within the sphere of influence of the black hole, while the other
  assumes Bondi-like accretion from the residual diffuse hot gas
  component. Each of the two models can produce $z\!=\!0$ galaxy
  populations similar both to that of our simple phenomenological
  model for `radio mode' feedback and to the observed population (see
  Fig.~\ref{physical}). Our main results are thus not sensitive to the
  details of the assumed accretion models.

\end{itemize}

The presence of heating from a central AGN has long been suspected as
the explanation for the apparent lack of gas condensation in cluster
cooling flows.  We have shown that including a simple treatment of
this process in galaxy formation models not only `solves' the cooling
flow problem, but also dramatically affects the properties of massive
galaxies, inducing a cut-off similar to that observed at the bright
end of the galaxy luminosity function, and bringing colours,
morphologies and stellar ages into much better agreement with
observation than is the case for models without such feedback. We will
extend the work presented here in a companion paper, where we
investigate the growth of supermassive black holes and the related AGN
activity as a function of host galaxy properties out to high
redshift. The catalogues of galaxies and supermasive black holes
produced by our modelling machinery are also being used for a very
wide range of projects related to understanding formation, evolution
and clustering processes, as well as for interpreting observational
samples.

\section*{Acknowledgements}
\label{acknowledgements}

DC acknowledges the financial support of the International Max Planck
Research School in Astrophysics Ph.D. fellowship.  GDL thanks the
Alexander von Humboldt Foundation, the Federal Ministry of Education
and Research, and the Programme for Investment in the Future (ZIP) of
the German Government for financial support.  Many thanks to Andrea
Merloni and Jiasheng Huang.  Special thanks to Eugene Churazov for
numerous valuable discussions and suggestions.  We would also like to
thank the anonymous referee for a number of valuable suggestions which
helped improve the quality of this paper.  The Millennium Run
simulation used in this paper was carried out by the Virgo
Supercomputing Consortium at the Computing Centre of the Max-Planck
Society in Garching.

\bibliographystyle{mnras}
\bibliography{./MF961rv}

\label{lastpage}

\end{document}